\documentclass[fleqn,usenatbib]{mnras}

\usepackage{newtxtext,newtxmath}
\usepackage[T1]{fontenc}

\DeclareRobustCommand{\VAN}[3]{#2}
\let\VANthebibliography\thebibliography
\def\thebibliography{\DeclareRobustCommand{\VAN}[3]{##3}\VANthebibliography}

\usepackage{graphicx}	% Including figure files
\usepackage{amsmath}
\usepackage{amsmath}
\usepackage{comment}
\usepackage{makecell}% Advanced maths commands
\usepackage{tablefootnote}
\usepackage{threeparttable}
\usepackage{booktabs}
\usepackage{siunitx}
\usepackage{lipsum}
\usepackage{multirow}
\usepackage{soul}

\title[ISM: Simulation and Observation]{Multiphase Neutral Interstellar Medium: Analyzing Simulation with \text{H~{\sc i}} 21cm Observational Data Analysis Techniques}

\author[Bhattacharjee et al.]{
Soumyadeep Bhattacharjee,$^{1,2}$ \thanks{E-mail: sbhatta2@caltech.edu}
Nirupam Roy,$^{1}$
Prateek Sharma,$^{1}$
Amit Seta$^{3}$
and Christoph Federrath$^{3,4}$
\\
$^{1}$Department of Physics, Indian Institute of Science, Bangalore, 560012, India\\
$^{2}$Department of Astronomy, California Institute of Technology, 1200E. California Blvd, Pasadena, CA, 91125, USA\\
$^{3}$Research School of Astronomy and Astrophysics, Australian National University, Canberra, ACT 2611, Australia\\
$^{4}$Australian Research Council Centre of Excellence in All Sky Astrophysics (ASTRO3D), Canberra, ACT 2611, Australia
}

\date{Accepted XXX. Received YYY; in original form ZZZ}

\pubyear{2015}

\begin{document}
\label{firstpage}
\pagerange{\pageref{firstpage}--\pageref{lastpage}}
\maketitle

\begin{abstract}

Several different methods are regularly used to infer the properties of the neutral interstellar medium (ISM) using atomic hydrogen (\text{H~{\sc i}}) 21cm absorption and emission spectra. In this work, we study various techniques used for inferring ISM gas phase properties, namely the correlation between brightness temperature and optical depth ($T_B(v)$, $\tau(v)$) at each channel velocity ($v$), and decomposition into Gaussian components, by creating mock spectra from a 3D magnetohydrodynamic simulation of a two-phase, turbulent ISM. We propose a physically motivated model to explain the $T_B(v)-\tau(v)$ distribution and relate the model parameters to properties like warm gas spin temperature and cold cloud length scales. Two methods based on Gaussian decomposition -- using only absorption spectra and both absorption and emission spectra -- are used to infer the column density distribution as a function of temperature. In observations, such analysis reveals the puzzle of large amounts (significantly higher than in simulations) of gas with temperature in the thermally unstable range of $\sim$200 K to $\sim$2000 K and a lack of the expected bimodal (two-phase) temperature distribution. We show that, in simulation, both methods are able to recover the actual gas distribution in the simulation till temperatures $\lesssim2500$~K (and the two-phase distribution in general) reasonably well. We find our results to be robust to a range of effects such as noise, varying emission beam size, and simulation resolution. This shows that the observational inferences are unlikely to be artifacts, thus highlighting a tension between observations and simulations. We discuss possible reasons for this tension and ways to resolve it.
\end{abstract}


%%%%%%%%%%%%%%%%%%%%%%%%%%%%%%%%%%%%%%%%%%%%%%%%%%

%%%%%%%%%%%%%%%%% BODY OF PAPER %%%%%%%%%%%%%%%%%%

\section{Introduction}\label{sec:introduction}

Understanding various properties, like temperature, density, velocity and magnetic fields, of the interstellar medium (ISM) provides valuable insights into the turbulent, multiphase ISM, formation and evolution of dense molecular clouds where stars form, and, consequently, the evolution of galaxies \citep{Ferriere01,Elmegreen04,MacLow04,McKee07,Padoan14,Federrath18, Ferriere2020, Griffiths23}. The atomic hydrogen (\text{H~{\sc i}}) is a major constituent of the neutral ISM, and thus the \text{H~{\sc i}} 21 cm emission and absorption spectra, arising from the transitions between the two hyperfine ground state levels, are among the most important tracers of the ISM \citep{Dickey90,Kalberla09,Griffiths23}. Since the first detections of \text{H~{\sc i}} spectrum from the ISM \citep{Ewen51,Muller51, Hagen55}, it has been used extensively in studying the thermodynamic, morphological, and turbulent properties of the ISM. 

The emission and absorption spectra are generally presented as the brightness temperature, $T_B(v)$, and optical depth, $\tau(v)$, in each velocity channel, $v$, respectively. The main physical parameters describing the ISM are the kinetic temperature, $T_k$, which quantifies the average thermal motion of the gas, spin temperature, $T_s$, which quantifies the hyperfine level population of neutral hydrogen, and column density, $N_{\text{H~{\sc i}}}$, which gives a measure of the amount of gas. Here we note that $T_s$ may not be equal to $T_k$, especially in the low-density warm phases in which collisions are insufficient \citep{Liszt01}. However, coupling with background Lyman-$\alpha$ radiation can bring the two temperatures closer through the Wouthuysen-Field effect  \citep[hereafter, WF effect;][]{Wouthuysen52,Field58}.

Thermal equilibrium models predict two stable phases of the neutral gas: cold neutral medium (CNM, $30 \mathrm{\ K}\lesssim T_k\lesssim 200 \mathrm{\ K}$) and warm neutral medium (WNM, $T_k\gtrsim 5000\mathrm{\ K}$), with the cold clouds being smaller and denser, existing as clumps within the large-scale warm medium \citep{Field69}. Contrary to this prediction, careful analyses of the \text{H~{\sc i}} spectra reveal that a significant fraction of the gas may lie in the ``unstable" phase with $200\mathrm{\ K}<T_k<5000\mathrm{\ K}$ \citep{Heiles03II, Kanekar03, Roy08, Roy13II, Murray15, Murray18}, the so-called unstable neutral medium (UNM). The most probable reason for the presence of unstable gas is turbulence originating from galactic spiral shocks, supernova explosions, and other physical processes \citep{Elmegreen04,Scalo04,Federrath17}. Turbulent heating and mixing may result in less cold and more unstable gas in the ISM. In spectral line observations, turbulence induces additional broadening of the lines, which makes temperature estimation of gas from line widths difficult.

There are various techniques for analyzing the observed \text{H~{\sc i}} absorption and emission spectra. The standard method is to decompose them into individual Gaussian components (termed Gaussian decomposition) with the underlying assumption that each component corresponds to an isothermal ISM cloud along the line of sight. Such Gaussian decomposition can give us the $T_s$ and $N_{\text{H~{\sc i}}}$ of the clouds corresponding to the Gaussian components. Recently, a method has been proposed that can decouple the thermal and turbulence broadening of these spectral components \citep[assuming pressure equilibrium and the velocity dispersion--size relation by][]{Larson81}, and gives us information about the kinetic temperature of the gas clouds \citep{KR19}. The absorption and emission spectra can also be directly used as a whole, without performing Gaussian decomposition, to estimate ISM properties like column density and cold gas fraction \citep{Heiles03II,Chengalur13,Roy13II,Murray18}. Although such analyses give us important information about the ISM, the reliability of the techniques and the underlying assumptions require further study and examination, which is the aim of this study.

Several numerical hydrodynamic or magnetohydrodynamic (MHD) simulations have been performed to understand the effect of turbulence on the thermal properties of ISM. Such simulations have been able to produce a significant amount of gas in the unstable phase through turbulence \citep{Audit05,Kim13,Kim14,Gazol16}. Many other properties of the ISM have been studied through simulations, such as the magnetic field morphology and the structure of the cold clouds and filaments \citep{Seta22,Beattie22,Lei22}. Besides such studies, simulations can also be used to test the observational techniques by creating synthetic \text{H~{\sc i}} absorption and emission spectra, applying the analysis techniques (e.g., Gaussian decomposition) to them, and testing the inferences against the ground truth from simulations.

Recently, a few studies have used simulation data to verify some of the widely used $\text{H~{\sc i}}$ $21$cm data analysis techniques. \citet{Kim14} used the hydrodynamic simulations by \citet{Kim13} to verify the methods for determining the line of sight column density, channel spin temperature, and cold gas fraction from the spectra. They also showed that the distribution of the spectral data points, $T_B(v)$ and $\tau(v)$, was qualitatively similar to the observed data by \citet{Roy13II}. \citet{Murray17} used the same simulation data to study the effectiveness of Gaussian decomposition in recovering the gas structures. They showed that the method could recover the temperature and column density of most structures within a factor of $\approx2$. More such studies, with different simulation data and analysis techniques, are needed to ascertain the validity of these techniques.

In this work, we undertake a systematic study of verifying several existing data analysis methods used frequently to infer the physical properties of the ISM from the observed spectra. We use the two-phase magnetohydrodynamic (MHD) simulation by \citet{Seta22} to create synthetic absorption and emission spectra and test the data analysis methodologies on them. Our work is broadly divided into two parts. The first part is dedicated to using the spectral lines as a whole without any Gaussian decomposition. These simple techniques can provide estimates of several ISM properties, avoiding the complexities involved with Gaussian decomposition. Here, we concentrate on the distribution of $T_B(v)$ and $\tau(v)$ and propose a physical and quantitative model to explain the observed distribution. In the second part, we perform a multi-Gaussian decomposition of the synthetic spectral lines and analyse the inferred properties. We verify the data analysis methodologies involving both the joint decomposition of absorption and emission spectra and using only the absorption spectra. We verify that the Gaussian decomposition method is largely capable of inferring the physical properties of the ISM, though we also discuss a few limitations related to the method. Please note that in comparing the properties inferred from the analysis methods to the actual physical properties of the simulation domain, we refer to the latter as the "true" properties. We clarify that the word should not be confused as indicating a true ISM property, which is yet to be established firmly.

The rest of the paper is organised as follows. In \S\ref{sec:simulations}, we briefly describe the main properties of the simulations and how we generate the synthetic observations of the simulations. In \S\ref{sec:spectral_lines_direct}, we use the data analysis methods without Gaussian decomposition and discuss a model for the $T_B(v)-\tau(v)$ distribution. In \S\ref{sec:gauss_decomp}, we describe the inferences from the Gaussian decomposition of absorption and emission spectra. We compare our results with previous works and observations in \S\ref{sec:discussions}, and discuss a few other aspects of the analysis techniques and some open questions. Finally, in \S\ref{sec:conclusion}, we summarise our results and conclude.

\section{Simulation and Synthetic Spectra}\label{sec:simulations}

\subsection{Simulation Data}

We use the recent two-phase MHD simulation of the ISM by \citet{Seta22}. The simulation uses the FLASH code to solve the non-ideal MHD equations in a triply periodic cubic numerical domain of $512$ grid points spanning $L=200$~pc along each axis, achieving a resolution of around $0.4$~pc (we have also tested our analysis methods with simulations of lower resolutions, see Appendix~\ref{app:resolution_effect}). It uses heating and cooling functions as prescribed in \citet{Koyama00, Koyama02, Vazquez-SemadeniEA2007}. For this work, we use the simulation with solenoidally driven turbulence, and we aim to extend our analysis in the future with other forms of driving \citep[compressive and a mix of both, see][for a discussion on different types of turbulence driving]{Federrath10}. The turbulence is driven at a length scale of half the simulation domain ($100$~pc) with velocity dispersion of $u_{\rm rms}=10\mathrm{\ km~s^{-1}}$. Explicit viscosity and resistivity have been added, yielding a hydrodynamic and magnetic fields Reynolds number ($u_{\rm rms}L/(2\nu)$ and $u_{\rm rms}L/(2\eta)$, where $\nu$ and $\eta$ are the viscosity and resistivity, respectively) of $2000$. The simulation does not use a galactic potential or any non-isotropic term in the set of MHD equations. Thus, the simulation domain is expected to be statistically isotropic. The simulation spans a period of $1$ Gyr, which is approximately $100$ times the eddy turnover time ($=(L/2)/u_{\rm rms}\approx 10$~Myr). For this work, we take the simulation at a time of $800$~Myr, by when the system has reached saturation in terms of its statistical properties and energies \citep[for further details, see figure~8 in][]{Seta22}. 

Unless otherwise mentioned, throughout the paper, we use the following definition for the gas phases: cold (CNM) - $T_k<200\mathrm{\ K}$, unstable (UNM) - $200\mathrm{\ K}<T_k<5000\mathrm{\ K}$, and warm (WNM) - $T_k>5000\mathrm{\ K}$. The simulation has $1.5\%$, $22.9\%$, $75.6\%$ by volume and $32.3\%$, $34.5\%$, $33.2\%$ by mass of cold, unstable, and warm gas, respectively. There is a very small amount of hot gas ($T_k>8000\mathrm{\ K}$) in the medium since the simulation does not include supernova explosions. Ionization of most warm gas is expected to be $<5\%$, with the unstable and cold gas ionization being even lower \citep{Wolfire95,Wolfire03,Liszt01}. Thus, in our analysis, we assume the gas in the simulation domain to be fully neutral. Here, we note that this may change depending mainly on the ionizing radiation field strength considered, which may result in a non-negligible ionization fraction, mostly for the warm phase. Such considerations are beyond the scope of this paper. However, we do not expect the results of this work to depend significantly on this effect other than the warm gas amount being appropriately scaled by the neutral fraction. Additionally, we also do not expect the small size of the simulation box ($200$ pc compared to several kilo-parsecs probed in observations) to affect the results as we are probing similar column densities of neutral gas with this simulation box as in observations ($10^{20}-10^{21}\mathrm{\ cm^{-2}}$), thus similar ranges of optical depth and brightness temperature in spectra. 

\subsection{Spectrum Generation}\label{subsec:spectrum_generation}

\begin{figure*}
    \centering
    \includegraphics[width=\linewidth]{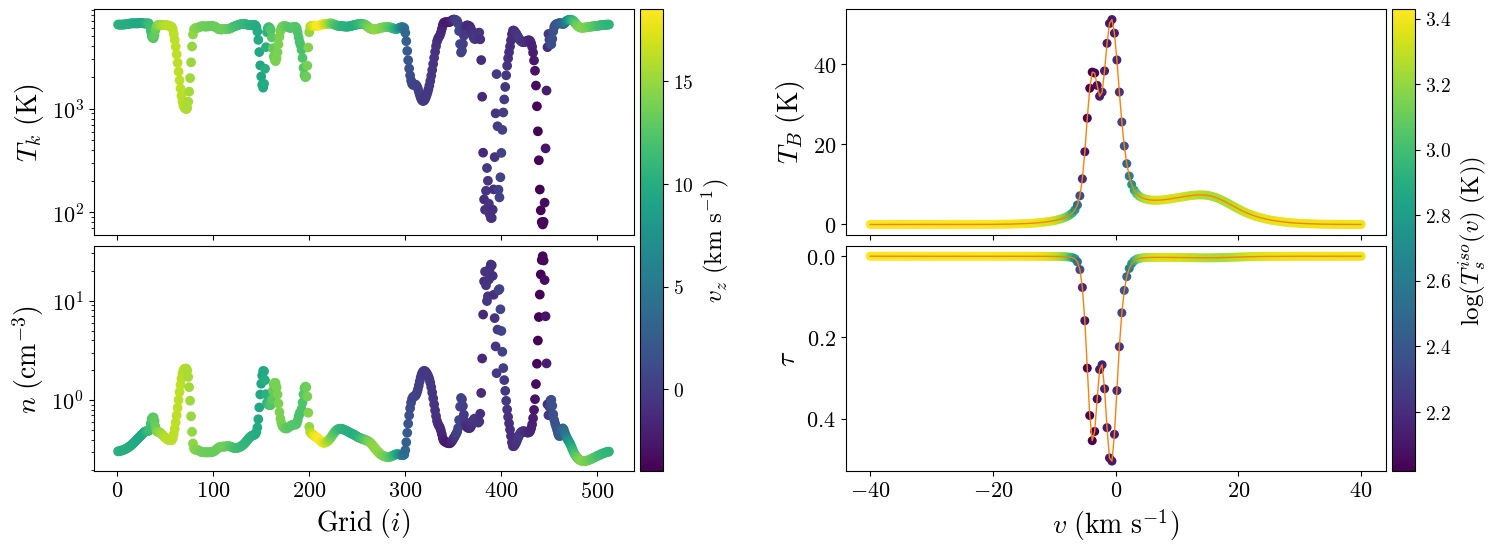}
    \caption{An example of a line of sight. \textbf{Left:} Variation of the kinetic temperature, $T_k$, and number density, $n$, along the line of sight, colored according to the line of sight velocity ($v_z$). The cooler clouds are usually associated with higher density. \textbf{Right:} The corresponding emission spectrum, $T_B(v)$, and absorption spectrum, $\tau(v)$, with the data points coloured according to $T_s^{\rm iso}(v)$ (Equation \ref{eq:ts_iso}). The appearance of a broad warm component in the emission spectrum and its absence in the absorption spectrum displays the inability of the absorption spectrum to detect warm gas.}
    \label{fig:example_spectrum}
\end{figure*}

The simulation results provide the values of the physical parameters, namely the temperature, density, and velocity for all the grid cells. However, in observations, the primary observables are only the absorption and emission spectra, which are a combined effect of all the gas along the line of sight. The former is obtained by recording the attenuation of the radiation from background point radio sources (like distant quasars), whereas the latter is obtained by recording the radiation intensity away from any bright background. Here, the simulation data is used to generate synthetic absorption and emission spectral lines. Due to the isotropic nature of the simulation, without loss of generality, we have taken the $z$ axis as the direction of a line of sight (LOS). We employ a synthetic spectra generation technique based on the solutions of the line-of-sight radiative transfer equation. We closely follow the formalism that has been applied in various previous works \citep{MD03,Kim14,Fukui18}. We describe the method employed below.

In generating the synthetic spectra, we consider each simulation grid cell to be a parcel of isothermal gas with the physical parameters given by the simulation results. For such a parcel, the absorption spectrum is dependent on the column density and spin temperature of the gas and additionally on the spectral profile of the gas, which is directly related to the velocity profile. For the $i^{\text{th}}$ grid, assuming a Gaussian velocity profile of each of the grid cells with mean velocity $u_{z,i}$ (the $z$ component of the velocity in the $i^{\text{th}}$ grid) and width $\sigma_i$, the optical depth, following the treatment for isothermal clouds in \citep{Draine11}, can be written as
\begin{equation}\label{eq:grid_od}
    \tau_i(v) = 2.190\frac{N_i}{10^{21}\mathrm{\ cm^{-2}}}\frac{100\mathrm{\ K}}{T_{s,i}}\frac{\mathrm{km~s^{-1}}}{\sigma_i}e^{-\frac{(v-u_{z,i})^2}{2\sigma_i^2}},
\end{equation}
where $N_i$ is the column density and $T_{s,i}$ is the spin temperature of the $i^{\text{th}}$ grid cell. $N_i = n_i\Delta z$, with $n_i$ being the number density of the $i^{\text{th}}$ grid cell and $\Delta z = 0.3906$ pc, the resolution of the simulation (the side length of a grid cell). Following the prescription in \cite{MD03}, the velocity profile width, $\sigma_i$, can be calculated as
\begin{equation}\label{eq:sigma_i}
    \sigma_i = \sqrt{\frac{k_BT_{k,i}}{m}+\left(\frac{du_{z,i}}{dz}\Delta z\right)^2},
\end{equation}
where $T_{k,i}$ is the temperature of the $i^{\text{th}}$ grid cell, $m$ is the hydrogen mass and 
\begin{equation}
    \frac{du_{z,i}}{dz}\Delta z \approx \frac{u_{z,i+1} - u_{z,i-1}}{2}.
\end{equation}
The first term inside the square root in Equation \ref{eq:sigma_i} is the thermal width, and the second term approximates the non-thermal (turbulent) broadening within the grid cell. Following the methodology in observations, for a bright enough background source with respect to which the self-radiation from the ISM can be neglected, the radiative transfer equation suggests that the observed optical depth will be a summation of the optical depths of all the gas parcels along the line of sight. Thus, the absorption spectrum is calculated as the optical depth spectrum
\begin{equation}
    \tau(v) = \sum_i \tau_i(v).
\end{equation}
The absorption spectra are dominated by the gas clouds with lower temperatures due to their lower spin temperatures and higher optical depths (following Equation \ref{eq:grid_od}).

The emission spectra are reported in terms of brightness temperatures, which are related to the radiation intensity by Rayleigh's blackbody relation \citep{Draine11}. To generate the spectrum, we use a ray-tracing approach and iteratively solve the following equation for the grid cells, $i$, along the line of sight as
\begin{equation}
    T_{B,i-1}(v) = T_{s,i}\left(1-e^{-\tau_i(v)}\right) + T_{B,i}(v)e^{-\tau_i(v)}.
\end{equation}
The first term in this expression is the absorption-corrected self-emission from the gas in the $i^{\mathrm{th}}$ grid cell; the second term represents the radiation from the grids behind, $T_{B,i}$, attenuated by the gas in the $i^{\text{th}}$ grid cell. Assuming that the $z=0$ plane faces the observer and there is no background source, we set $T_{B,512}(v) = 0$ (512 being the total number of grid cells in each direction of the cubic simulation domain). This makes $T_{B,0}(v)=T_B(v)$ the brightness temperature spectrum seen by the observer. 

Figure \ref{fig:example_spectrum} shows an example of synthetic absorption and emission spectra, along with the temperature and density variation along the line of sight. In the optical depth spectrum, the cooler gas along the line of sight manifests as distinct narrow and high-amplitude components. The broader warm components are subdominant owing to their high spin temperature (see Eq. \ref{eq:grid_od}) and low optical depth. The latter, however, are clearly present in the emission spectrum.

We note here that, as also mentioned in \S\ref{sec:introduction}, the kinetic temperature $T_k$ and spin temperature $T_s$ may not be the same, especially at higher temperatures at which collisions are not sufficient to equilibrate the two temperatures. Alongside several other processes, the WF effect plays an important role in determining the relation between the two temperatures. We present our results assuming a constant but inefficient WF effect resulting in $T_s<T_k$. We use the numerical results of \citet{Liszt01} to relate the two temperatures. However, several recent studies indicate that the WF effect in the ISM maybe efficient enough to render the two temperatures equal for all phases \citep{Murray17,Seon20}. Thus, we have also performed our analyses using $T_s=T_k$ for all phases. We discuss the results for this case separately in \S\ref{subsec:mwf_discussion}.  

We also note that observed ISM spectra are contaminated by noise, which introduces additional complexities in analyzing the data. An additional limitation with observations is the larger beam size associated with the emission spectra compared to the absorption spectra. We separately incorporate both of these effects in our study (see Appendix \ref{appendix:noise_effect} and \ref{appendix:emission_beam_effect}, respectively). For simplicity, we primarily present our analysis without these effects, but we refer to the results with these effects when required.

\section{Analysis Without Gaussian Decomposition}\label{sec:spectral_lines_direct}

\subsection{Estimating total column densities}

The emission and absorption spectra can be used directly to extract several physical properties of the ISM, namely the total column density and cold gas fraction. An unbiased\footnote{\textcolor{black}{For a sufficiently large sample size, the ratio of the estimated to the true quantity (column density in the present case) shows a roughly equal distribution above and below unity (see the left panel of Fig. 6 in \citealt{Kim14}).}} estimator of column density, the isothermal estimator, is given by \citep{Dickey82, Chengalur13}
\begin{equation}\label{eq:nh1iso}
\begin{split}
    N_{\text{H~{\sc i}}}^{\rm \rm iso}/10^{21} = 1.823\times10^{-3}\mathrm{\ cm^{-2}\ }\int T_s^{\rm \rm iso}(v)\tau(v) dv, \\ = 1.823\times10^{-3}\mathrm{\ cm^{-2}\ }\int \frac{T_B(v)\tau(v)}{1-e^{-\tau(v)}}dv.
    \end{split}
\end{equation}
This is an isothermal estimator of the column density, with 
\begin{equation}\label{eq:ts_iso}
    T_s^{\rm \rm iso}(v) = \frac{T_B(v)}{1-e^{-\tau(v)}}
\end{equation}
being the isothermal estimator of the spin temperature at the velocity channel $v$ (the points in the right panel of Fig. \ref{fig:example_spectrum} are coloured according to $T_s^{\rm iso}[v]$). These estimators assume that the radiation in a single velocity channel arises from an isothermal gas with temperature $T_s^{\rm \rm iso}(v)$. $T_s^{\rm \rm iso}(v)$, additionally, turns out to be an unbiased estimator of the optical-depth-weighted spin temperature $T_s^{\rm avg}(v)=\sum_i T_{s,i}\tau_i(v)/\sum_i\tau_i(v)$ of the gas along the line of sight. We verify that $N_{\text{H~{\sc i}}}^{\rm \rm iso}$ agrees with the true column density within a factor of $\approx1.2$ and $T_s^{\rm \rm iso}$ agrees with $T_s^{\rm avg}$ within a factor of $\approx1.5$. 

The absorption and emission spectra can be used to determine the cold gas content along the line of sight. The cold gas fraction can be estimated using a parametric cold gas spin temperature, $T_c$, and the line of sight average spin temperature $T_s^{\rm \rm los} \left(= \int \tau(v) T_s^{\rm \rm iso}(v) dv/\int \tau(v) dv\right)$ as $f_c = T_c/T_s^{\rm \rm los}$ \citep{Dickey09,Kim14}. We found that $T_c$ values in the range of $50-100\mathrm{\ K}$ fit well the cold gas fraction for most of the sightlines. These results are consistent with those obtained in \citet[see figs.~5 and~6 in][]{Kim14}.

\subsection{$T_B(v)$--$\tau(v)$ space: A new model}\label{subsec:tb_tau_model}
The optical depth and brightness temperature in a velocity channel are impacted by both the spatial and velocity distribution of the various phases of gas along the line of sight. The distribution of the spectral data points, $T_B(v)$ and $\tau(v)$, coupled with the channel spin temperature $T_s^{\rm \rm iso}(v)$, can provide valuable information about the properties of the ISM and has been used in several recent works \citep{Roy13,Kim14,Saha18,Basu22}. There have been previous attempts to explain the $T_B(v)-\tau(v)$ distribution using two-phase ISM models \citep{Strasser04} or using simulations \citep{Kim14}. Although the general properties of this space could be understood, the analysis was restricted to mostly a qualitative understanding. Here we propose a quantitative model based on physical arguments to describe the boundaries of the distribution of data points in the $T_B(v)$--$\tau(v)$ space. We apply our model to the synthetic spectral data generated from our simulations and relate the model parameters to the physical properties of the simulation domain. To test the applicability of the model to real data, we take the Giant Meterwave Radio Telescope(GMRT)/Westerbork Synthesis Radio Telescope (WSRT)/Australia Telescope Compact Array (ATCA) observations of \citet[][37~lines of sight]{Roy13,Roy13II} and fit our model. 

\begin{figure*}
    \centering
    \includegraphics[width=\linewidth]{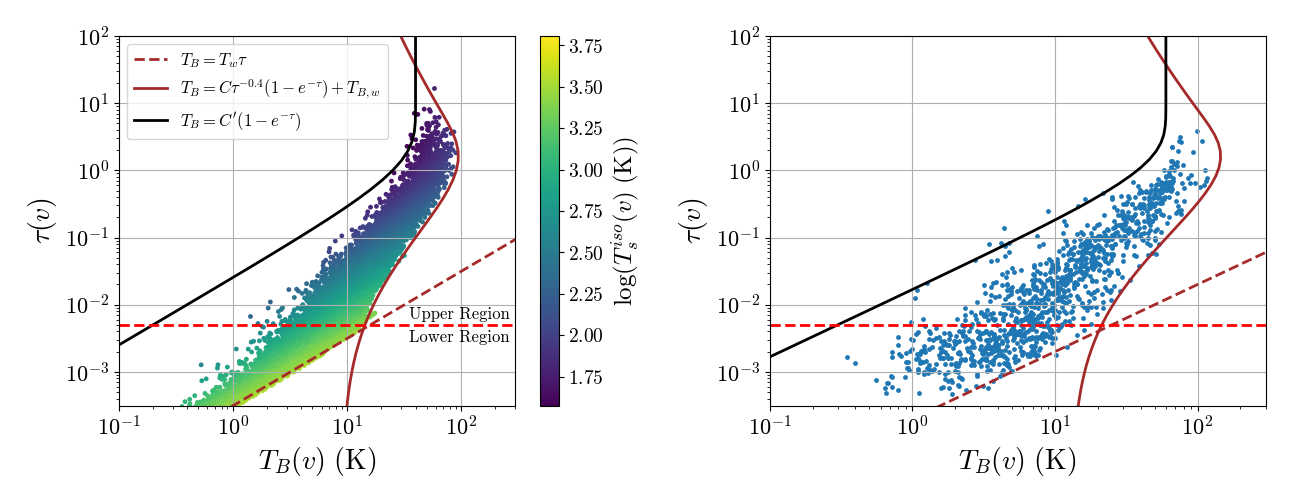}
    \caption{The $T_B(v)$--$\tau(v)$ distribution for synthetic spectra for randomly chosen $1000$ lines of sight (\textbf{Left}, the points are coloured according to $T_s^{\rm iso}(v)$, Equation \ref{eq:ts_iso}) and observed data from \citet{Roy13} (\textbf{Right}). The red horizontal line marks $\tau=0.005$. The brown lines denote the models for the right boundary (see section \ref{subsubsec:right_tbtau_model}). Both the simulation and observational data are well-fitted by the model. The parameters for synthetic and observed data are: $(T_W,\ C,\ T_{B,w}) = (3200\mathrm{\ K}, \ 130\mathrm{\ K}, \ 9\mathrm{\ K})\text{ and }(5000\mathrm{\ K}, \ 200\mathrm{\ K}, \ 13\mathrm{\ K})$ respectively. The black solid line shows the left boundary model (see section \ref{subsubsec:left_tbtau_models}) which fails to fit the synthetic data satisfactorily. 
    }
    \label{fig:tb_tau_inner_both}
\end{figure*}

Figure \ref{fig:tb_tau_inner_both} shows the $T_B(v)$--$\tau(v)$ space for the synthetic spectra (left) and observed data (right), with our models (described in \S\ref{subsubsec:right_tbtau_model} and \S\ref{subsubsec:left_tbtau_models})  for the boundaries in the $T_B$--$\tau$ space overplotted. The data points are contained within a band in this space; the right boundary is relatively more prominent than the left. $T_B(v)$ is also seen to increase with optical depth up to a certain value, post which absorption dominates and the value decreases. Our models aim to relate these trends to the quantitative physical properties of the neutral ISM.

\subsubsection{The ``non-warm'' gas spin temperature and optical depth}\label{subsubsec:tb_tau_model_derivation}

Before going into the models, we set up the ground by deriving a relationship between the channel spin temperature of the non-warm (cold or unstable) ISM gas, $T_{nw}(v)$, to the optical depth, $\tau(v)$. We shall use the following formalism to understand the right boundary of the $T_B(v)$--$\tau(v)$ distribution. We start with the optical depth equation for an isothermal gas cloud,
\begin{equation}
\label{eq:tau}
\tau(v) = K\frac{N_{\text{H~{\sc i}}}}{T_{nw}\sigma}e^{-\frac{(u-v)^2}{2\sigma^2}},
\end{equation}
which is the same as Equation \ref{eq:grid_od}, except not for a grid cell, with $K$ being a product of numerical constants. For non-warm gas, we can approximate the spin and kinetic temperatures to be equal, which is reasonably valid till $\approx2000\mathrm{\ K}$ for any strength of the WF effect. We assume an isobaric equation of state with pressure $P$ \citep[simulations show that lower-temperature gas more closely follows an isobaric equation of state, see, for example, fig.~3 in][]{Seta22}. We further assume that $\sigma \sim \sqrt{~T_{nw}}$ as the extent of turbulent broadening in smaller-scale non-warm gas will not affect our formalism \citep[turbulent broadening is directly related to the gas length scales, see][and Equations \ref{eq:larson_scaling} and \ref{eq:sigma_turb_sim} for the scaling relations]{Larson79,Larson81}. Using these simplifications, we can write
\begin{equation}\label{eq:tnw_equation}
\begin{split}
    \tau(v) = K\frac{P\ell}{T_{nw}^{2.5}}e^{-\frac{(u-v)^2}{2\sigma^2}}, {\rm~ or}\\ %\longrightarrow \\ 
    T_{nw} = (KPl)^{0.4}\tau(v)^{-0.4}e^{-0.2\frac{(u-v)^2}{\sigma^2}},
    \end{split}
\end{equation}
where $\ell$ represents the typical line-of-sight length scales of the non-warm clouds. For a given $\tau(v)$, $T_{nw}$ is the maximum for $u = v$. This gives the maximum possible temperature of the non-warm cloud, which may yield the given $\tau(v)$. We further note that non-warm gas in the ISM is not expected to have arbitrarily large length scales. Consequently, the pre-factor attains a maximum at about the largest non-warm length scale value for a given $P$. We define the parameter $C=(KPl)^{0.4}$. Owing to the weak exponent, small variations in $l$ or $P$ will not affect the parameter significantly. This leads us to the following expression:
\begin{equation}\label{eq:max_tnw}
    T_{nw}(v) = C\tau(v)^{-0.4}.
\end{equation}
This equation represents the maximum temperature of an isothermal non-warm gas cloud which may result in the optical depth given by the spectrum at a particular velocity channel. We recognize that a spectrum is composed of the optical depth profiles of all the different gas clouds along the line of sight. The temperature calculated using Equation \ref{eq:max_tnw} is only an isothermal upper limit estimated from the optical depth at a velocity channel. We note here that a similar relationship of temperature to optical depth has also been reported by \citet{Saha18} and is also similar to the classical $T_s - \tau$ relation \citep{Kulkarni88,Heiles03II}. The above re-parameterization has naturally led to such a relation.

\subsubsection{The right envelope of the $T_B(v)-\tau(v)$ distribution}\label{subsubsec:right_tbtau_model}

In this section, we discuss the model for the right envelope and its implications. We divide the $T_B(v)-\tau(v)$ relation broadly into two regions: the lower region, which is characterized by very low optical depths ($\tau\lesssim 0.005$, the lower region), and the rest (upper region). In the lower region, the spectra are expected to be dominated by the warm gas, either from the wings of their broad spectral components or from the velocity channels with no cold gas contributions. This is also supported by the high $T_s^{\rm iso}(v)$ value in the region (colourbar in Figure \ref{fig:tb_tau_inner_both}, left plot). For this region, our model assumes a simple form given by
\begin{equation}
\label{eq:tb_warm}
    T_B(v) = T_w(1-e^{-\tau(v)})\approx T_w\tau(v),
\end{equation}
where $T_w$ is a parameter of our model and represents the typical spin temperature of the warm gas in the ISM. The final approximation is valid in the low optical depth regime. 

The upper region ($\tau\gtrsim 0.005$), characterized by higher optical depths, is dominated by non-warm gas components. We assume that the non-warm component is embedded within the warm component, with a fraction $q$ of the warm component being in front of the non-warm component. Assuming negligible optical depth from the warm components, the brightness temperature then becomes
\begin{equation}\label{eq:tb_tau_equation}
    T_B(v) \approx T_{nw}(1-e^{-\tau(v)}) + T_{B,w}q + T_{B,w}(1-q)e^{-\tau(v)},
\end{equation}
where $T_{B,w}$ is the parametric brightness temperature of the total warm gas column and $T_{nw}$ is the parametric spin temperature of the non-warm gas. The first term in the equation is the self-emission term from the non-warm gas. The second and third terms represent the emission from the warm gas in front and behind the non-warm cloud, respectively. The true value of $q$ varies with both line-of-sight and velocity channels and is difficult to determine. In this model, however, $q$ appears as an effective parameter. Previous studies have shown that a single set of $q$, $T_{nw}$ and, $T_{B,w}$ values cannot satisfactorily explain the complete envelope \citep{Strasser04} and thus the parameters must vary over the envelope. As optical depth does not depend on the relative positions of the clouds along the line of sight, $q$ cannot be a function of $\tau$. Rather, following the formalism in \S\ref{subsubsec:tb_tau_model_derivation}, we argue that $T_{nw}$ is a function of the optical depth. We note that we can maximize $T_B(v)$ by setting $q=1$, which physically means that the right boundary is primarily constituted by velocity channels where the non-warm component lies behind the warm component. We further maximize $T_B(v)$ by maximizing $T_{nw}$ using Equation \ref{eq:max_tnw}. Thus, we arrive at the model for the upper region of the $T_B(v) - \tau(v)$ space:
\begin{equation}\label{eq:tb_tau_model_eq2}
    T_B(v) = C\tau(v)^{-0.4}\left(1-e^{-\tau(v)}\right) + T_{B,w},
\end{equation}
where $C$ and $T_{B,w}$ are parameters of our model. 

\begin{table}
\centering
\caption{The parameter values for the model of the right boundary of the $T_b(v)-\tau(v)$ distribution for the different cases assumed in this work. cWF and mWF represent constant and maximum WF effect cases, respectively. The parameters $C$ and $T_{B,w}$ do not change significantly between the two cases; thus, only one value is reported.}
\label{tab:tb_tau_model_params}
\begin{threeparttable}
\resizebox{\linewidth}{!}{%
\begin{tabular}{|c|c|c|c|c|} 
\hline
\multirow{2}{*}{} & \multirow{2}{*}{$C$} & \multirow{2}{*}{$T_{B,w}$} & \multicolumn{2}{c|}{$T_w$}    \\ 
\cline{4-5}
                  &                    &                      & cWF  & mWF                 \\ 
\hline
Noiseless         & 130                & 10                   & 3200 & 6000                \\
With Noise$^\text{a}$        & 130                  & 10                    & N/A$^\text{c}$    & N/A$^\text{c}$                   \\
$N=3^\text{b}$               & 110                & 12                   & 3400 & 6000                \\
$N=5^\text{b}$               & 100                & 18                   & 4500 & 7800                \\
$N=15^\text{b}$               & 60                & 25                   & 8000 & 15000                \\
\hline
GMRT/WSRT/ATCA    & 200                & 13                   & \multicolumn{2}{c|}{5000}  \\
\hline
\end{tabular}
}
\begin{tablenotes}
       \item [a] See Appendix \ref{appendix:noise_effect}
       \item [b] See Appendix \ref{appendix:emission_beam_effect}
       \item [c] Difficult to estimate with the assumed noise levels
       
     \end{tablenotes}
\end{threeparttable}
\end{table}

Figure \ref{fig:tb_tau_inner_both} (left) shows the model fit\footnote{We performed a fit-by-eye as numerical fitting of the boundary is difficult. All the quoted parameter values may be taken with an error of $5-10\%$} for the synthetic spectra (brown lines) for the constant WF effect case. We see that the model resembles the right envelope quite well with the parameter values $T_w = 3200\mathrm{\ K}$, $C=130\mathrm{\ K}$, and $T_{B,w}=10\mathrm{\ K}$. We note that with a constant but inefficient WF effect and following the \citet{Liszt01} relation between $T_s$ and $T_k$,  $T_s$ is mostly limited to $\approx 3000\mathrm{\ K}$ even for the warm gas with much higher kinetic temperatures. Thus, $T_w$ gives a good estimate of the warm gas spin temperature. The $T_{B,w}$ value well represents the typical brightness temperature of the warm components. Using the $T_{B,w}$ and $T_{w}$ parameters, we can estimate the typical optical depth from the warm gas components as $\tau_w \approx T_{B,w}/T_{w} \approx 0.003$ (see Eq. \ref{eq:tb_tau_equation}), which agrees well with the spectra. The $C\tau^{-0.4}$ factor represents the temperature of the non-warm component, which dominates at a given optical depth. The transition from the lower region to the upper region in the $T_B-\tau$ plane occurs at around $\tau \approx 0.005$. At this optical depth, the temperature with $C=130\mathrm{\ K}$ is $\approx 1100\mathrm{\ K}$ (see Eq. \ref{eq:max_tnw}), which lies in the unstable phase. At $\tau=1$, the temperature is $130\mathrm{\ K}$, which lies in the cold phase. Thus, we see that gas with a wide range of temperatures contributes to the $T_B(v)-\tau(v)$ distribution. Table \ref{tab:tb_tau_model_params} summarizes the model parameters for all the different cases considered in this section. 

The right panel in Figure \ref{fig:tb_tau_inner_both} (right) shows our model fit for the observed data. The model parameters are: $T_w = 5000\mathrm{\ K}$, $C=200\mathrm{\ K}$, $T_{B,w}=13\mathrm{\ K}$. The parameters $T_w$ and $C$ are significantly different than those for the synthetic data. This brings out the quantitative difference between the simulation and observed data. The high $T_w$ for the observational data implies the existence of a significant amount of gas with a high spin temperature (which may indicate an efficient WF effect in the ISM; see \S\ref{subsec:mwf_discussion} for further discussion). Considering the parameter $C$, we see that at the transition from the lower to the upper region ($\tau\approx0.005$), the temperature of the non-warm component is $\approx 1800\mathrm{\ K}$ (see Eq. \ref{eq:max_tnw}), which is much higher than for the simulation and lies in the unstable range. We relate this to the detection of a significant amount of gas in the temperature range of $1000-2000\mathrm{\ K}$, the unstable gas, in several studies \citep{Heiles03II,Roy13,KR19}. However, we note here that practical limitations with observations like larger emission beam size or noise may also affect the parameter values and may lead to larger uncertainties (see Table \ref{tab:tb_tau_model_params} and Appendix \ref{appendix:noise_effect} and \ref{appendix:emission_beam_effect}). For consistency checks and to constrain the parameters better, it is important to perform the analysis with emission spectra with narrower emission beam widths (e.g., using HI4PI data). However, for this work, we stick to the data from \citet{Roy13} and defer further analyses for the future (see Appendix~\ref{app:lab_hi4pi}).

Assuming $T_{B,w}$ to be representative of the peak brightness temperature of the warm gas and that the line-width $\sim\sqrt{~T_{k,w}}$, where $T_{k,w}$ is a parametric warm gas kinetic temperature, we can obtain an approximate relation between $T_{B,w}$ and the warm column density: $N_{\text{H~{\sc i}}}^w/\left(10^{21}\mathrm{\ cm^{-2}}\right)\approx T_{B,w}\sqrt{~T_{k,w}}/\left(2409\mathrm{\ K^{3/2}}\right)$ (combining Eqs. \ref{eq:tau} \& \ref{eq:tb_warm}). The $T_{B,w}$ values for the synthetic and observed data and typical values of $T_{k,w}$ yield $N_{\text{H~{\sc i}}}^w\approx\mathrm{few\times}10^{20}\mathrm{\ cm^{-2}}$. This well represents the warm gas column density of the simulation and verifies the warm gas origin of the parameter $T_{B,w}$. For the ISM, this is comparable to typical warm gas column densities \citep[and tallies with the minimum warm column density limit for CNM clouds to form; see][]{Kanekar11}.

From the derivation in \S\ref{subsubsec:tb_tau_model_derivation}, we see that the parameter $C$ is related to the characteristic maximum length scales of the non-warm components. Thus, given a value of the parameter $C$, the relation can be inverted to obtain an estimate of the maximum limit of the non-warm length scale. Substituting the constants, we find $\ell = C^{2.5} / 7.434P \mathrm{\ pc}$, where $P$ is the pressure of the non-warm gas. Using the volume-averaged pressure $P\approx2717 \mathrm{\ K~cm^{-3}}$ for the simulation, we estimate a length scale of $\approx9$~pc. Such a length scale is associated with the larger non-warm clouds in the simulation domain (the non-warm length scale in the simulation varies from $\sim 0.8$ to $\sim20$ parsecs). A similar length scale estimation for the observed data yields a value of $\approx 20$ pc when the pressure is taken to be $3700 \mathrm{\ K/cm^3}$ \citep[nominal ISM pressure][]{McKee77,McKee95,Jenkins11,KR19}. This is close to the cold cloud scales measured from Zeeman splitting \citep[$\sim10\mathrm{\ pc}$,][]{Heiles05}, and happens to be close to the cooling length \citep[$\sim20\mathrm{\ pc}$,][]{Hennebelle99}.

\subsubsection{The left envelope of the $T_B(v)-\tau(v)$ distribution}\label{subsubsec:left_tbtau_models}
Unlike the right envelope, the left envelope is not very well defined. This region of the $T_B(v)-\tau(v)$ space mostly comprises cold clouds with small length scales or the outer wings of multiple cold Gaussians \citep[also see \S4.1 in][for a related discussion]{Kim14}. The contribution from the warm component in this regime is subdominant. This leaves us with the model $T_B(v) = T_{nw}\left(1-e^{-\tau(v)}\right)$ (see Eq. \ref{eq:tb_tau_equation}). The left region corresponds to the lowest $T_B$ points for a given $\tau$, and demands minimization of $T_{nw}$. Unlike maximization, Equation \ref{eq:tnw_equation} does not yield a non-trivial minimum limit for $T_{nw}$ for a given $\tau$. Rather, in this case, the limit arises from the lowest possible gas temperature in ISM. We see that a constant value of $T_{nw} = 60\mathrm{\ K}$ can enclose most of the data points of the observed data (black line in Figure \ref{fig:tb_tau_inner_both}, right panel). Such a model was also proposed earlier with a similar cold gas temperature value by \citet{Strasser04}. It is interesting to note that this model intersects with the previous model for the right envelope, which puts a limit on the optical depth and brightness temperatures. However, such a model (with $C'=40~\mathrm{\ K}$, black line in the left panel of the same figure) cannot explain the left envelope of the synthetic data satisfactorily. 

The data points forming the left envelope, as noted in \citet{Kim14}, are expected to arise from sub-parsec scale cold clouds or the Gaussian tails of the cold components. Considering the former, in the ISM, the perceived small length scales of cold clouds can arise from the large asymmetry in the structure of cold gas, which may exist as sheets with aspect ratios as high as $\sim200$ \citep{Heiles05}. If this is the main reason, the discrepancy should decrease with increasing simulation resolution. Contrary to this expectation, our resolution study shows a different trend (fewer data points in the left region with increasing resolution, see Appendix~\ref{app:resolution_effect}). However, there is a possibility that simulations with much higher resolution than this work are needed to resolve these structures (e.g.~sheets) and see their prominence in the $T_B(v)-\tau(v)$ distribution.

For the other cause, the tails of the narrow Gaussians of the cold clouds usually get buried under the broad, warm components of the surrounding gas. Thus, they do not surface in the $T_B(v)-\tau(v)$ distribution. However, cold gas with a velocity significantly different than the surrounding gas may result in a cold Gaussian in the spectrum separated from the rest of the gas. The wings of such components may contribute to the left boundary of the distribution. This occurs in the lowest resolution simulation ($128^3$) due to a sudden jump in the velocity and temperature (see Figure~\ref{fig:128_problem_exs} in Appendix~\ref{app:resolution_effect}). However, a similar effect may occur in the ISM with fast-moving cold components through the surrounding (relatively warmer) medium, similar to intermediate and high-velocity clouds \citep[IVCs and HVCs,][]{Wakker97,EliseAlbert2005}, or supernova shock fronts.

A different origin of the leftmost data points may be the larger emission beam width. A larger beam results in an averaging of the emission spectra over a region in the plane of the sky. If this length scale becomes comparable to the length scale of cold clouds, the resultant brightness temperature of these clouds decreases. Absorption spectra, on the other hand, are obtained over a narrow beam (as the background sources are usually point-like). This may result in an apparent lower brightness temperature for a given optical depth (equivalent to being in the left region of the $T_B(v)-\tau(v)$ distribution) for the cold components. Our study of this effect (detailed in Appendix~\ref{appendix:emission_beam_effect}) indeed shows that with increasing emission beam width, the left boundary shifts towards left (a decreasing value of the parameter $C'$) with increasing occurrences of such low $T_B(v)$ points in the left region (see the extreme case of $N=15$ in Figure~\ref{fig:N_3_5_tb_tau}).

\section{Multi-Gaussian Decomposition}\label{sec:gauss_decomp}

\begin{figure*}
    \centering
    \includegraphics[width=\linewidth]{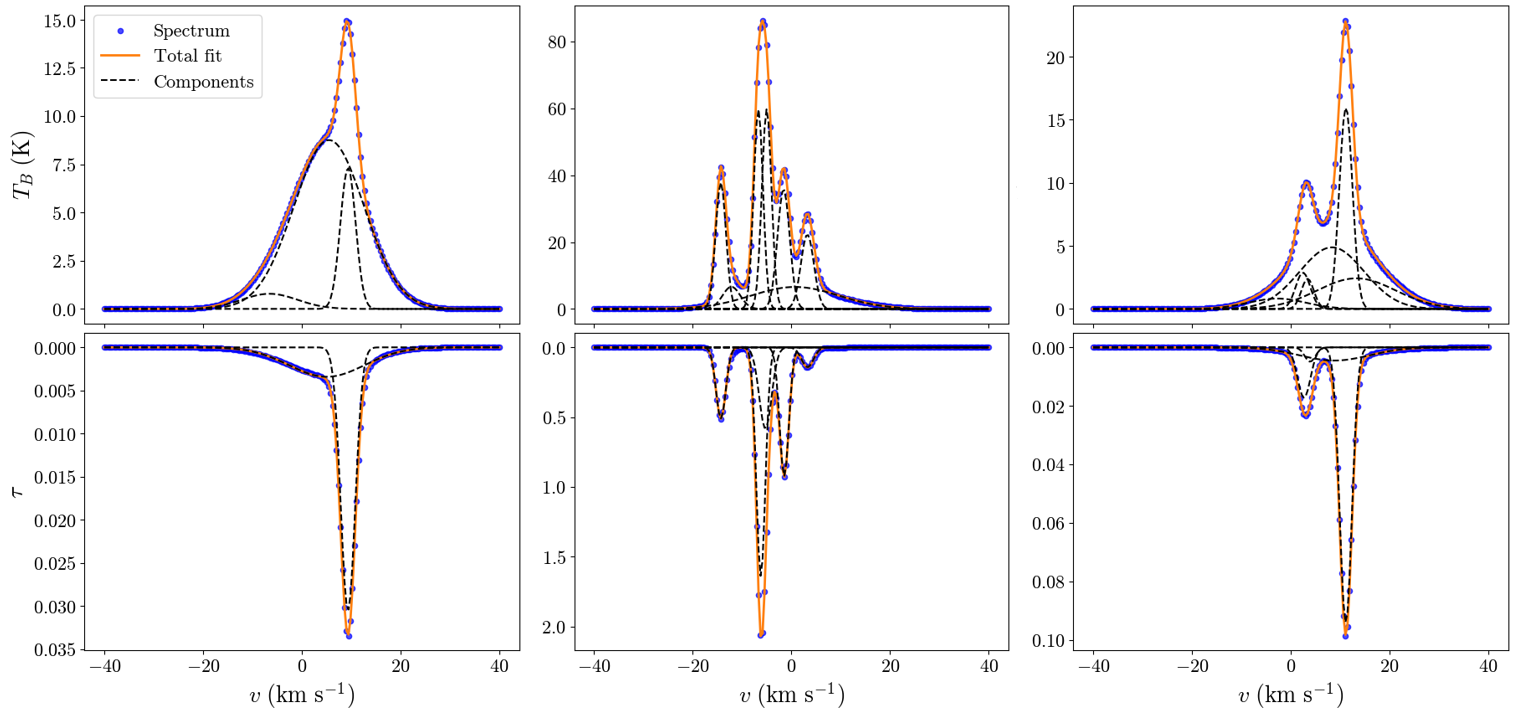}
    \caption{Examples of absorption and emission spectra and their Gaussian decomposition results. The blue points are the spectral data points, and the orange line represents the net Gaussian fit (sum of all Gaussian components). The black dashed lines show the individual Gaussian components inferred. Several warm components subdominant in absorption shows up as extra components in emission (the emission-only components). On an average, there are about $2-4$ absorption-emission joint Gaussian components and $0-2$ emission-only components per line of sight.}
    \label{fig:em_abs_gauss_decomp_exs}
\end{figure*}

Gaussian decomposition is a widely used technique to extract information about isothermal gas clouds along the line of sight from the absorption and emission spectra. Here we recognize that emission spectra for the cloud components may not be Gaussians in optically thicker sightlines (since $T_B \propto \tau$ only for $\tau \lesssim 1$), which is the case for several of our synthetic spectra. Nevertheless, the emission spectrum can be well decomposed into Gaussians and can be used to reasonably estimate the spin temperature and column density of the components. Such methods have been used in analyzing data from several surveys like LAB \citep{Haud07,Kalberla18, Marchal21} and other studies with synthetic spectra \citep{Murray17}. A theoretically-motivated method that accounts for two phases, which was used in several later works \citep[SPONGE and MACH surveys, see][]{Murray15,Murray18,Nguyen19,Murray21}, is outlined in \citet{Heiles03}. However, this method suffers from the uncertainty related to the relative position of the gas clouds along the line of sight. For the purpose of this work, we use the former method.

We consider two analysis methods involving multi-Gaussian decomposition of the spectra. First is the standard method of performing joint Gaussian decomposition (JGD) of the absorption and emission spectra. The joint components are used to find the spin temperature and column density of the corresponding gas cloud. However, the emission components suffer from several drawbacks, mostly related to the much broader beam width associated with them than the absorption spectrum. This leads to stray emissions contaminating the spectrum. To overcome such problems, \citet{KR19} proposed a method of using the Gaussian components from only the absorption spectra to determine the component properties (henceforth, the KR method). We use this method in \S\ref{subsec:using_only abs}.

Recently, several automatic Gaussian decomposition algorithms, which can identify the number of components and fit the spectra, have been developed \citep{Lindner15,Marchal19}. However, such algorithms have their own uncertainties, and they also fail to work on noiseless data. Additionally, they cannot be directly used for joint fitting of absorption and emission spectra. In this work, we use our own algorithm for automatic Gaussian decomposition, which is suitable for all the different cases considered in this work. 

Our algorithm uses \textsc{lmfit} \citep{lmfit23} and is based on repetitive fitting with an increasing number of components (see Appendix \ref{appendix:gauss_decomp_algo} for the details). For the joint absorption-emission decomposition, we have restricted (within a small window) the component line centers and width in emission with those obtained from fitting the absorption spectrum. Our method ensures that for each absorption component, we get a corresponding emission component (the joint absorption-emission components). We also get several extra components from the emission spectra, which do not appear in absorption. These components are mostly from the warm gas, which is subdominant in absorption, but detected in emission. We refer to these as ``emission-only" components. We have also set several criteria (see Appendix \ref{appendix:filtering_comps}) for accepting the inferred components. Due to the fundamental difference between the two methods, the criteria for the JGD and KR methods are different. For the first, the criteria involve the joint components, whereas, for the latter, the criteria are based on only the absorption components (see Appendix \ref{appendix:filtering_comps} for the details). Thus, the number of components used in the two methods is different. We randomly choose $200$ sightlines through the simulation domain for our analysis. In Figure \ref{fig:em_abs_gauss_decomp_exs}, we provide a few examples of the Gaussian decomposition of both the emission and absorption spectra. The synthetic spectra, the multi-Gaussian fit to the data, and the individual inferred components are shown.

\subsection{Properties of the Gaussian components}

\begin{figure*}
    \centering
    \includegraphics[width=\linewidth]{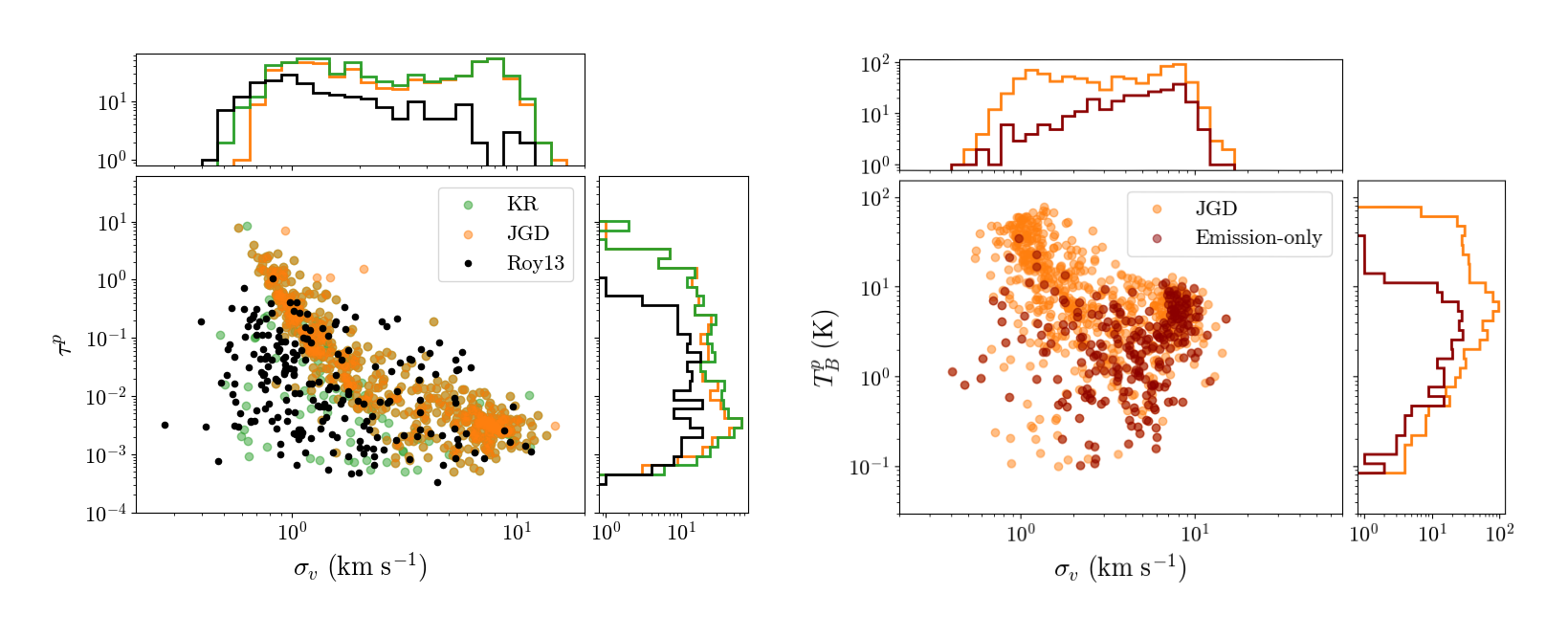}
    \caption{\textbf{Left:} The width, $\sigma_v$, and amplitude, $\tau^p$, of the absorption components from the KR method (green points) and those from JGD (orange points). For comparison, the absorption data from \citet{Roy13} are also plotted (black points). The observed components do not follow the distribution of the synthetic components, showing a  quantitative difference between the observations and the simulation. (\S\ref{subsec:comparing_prev_works_and_obs}). \textbf{Right:} The peak brightness temperature, $T_B^p$, and $\sigma_v$ of the emission components from JGD and the emission-only components. As expected, warm gas with larger $T_B^p$ and $\sigma_v$ is missing in the emission-only data. The histograms show the distribution of the parameters corresponding to the respective axes for both plots.}
    \label{fig:abs_em_comps_scatter_dist_cwf_roy}
\end{figure*}

With noiseless spectra of 200 random sightlines through the numerical domain, JGD yields a total of $532$ joint absorption-emission components and $263$ emission-only components. For the KR method, we obtain $604$ absorption components. The components are characterized by the peaks $T_{B}^p$ and $\tau^p$ for emission and absorption components, respectively, and the total line broadening $\sigma_v$. Table \ref{tab:gas_fractions} summarizes the Gaussian decomposition results for all the different cases (including various observational studies). 

Figure \ref{fig:abs_em_comps_scatter_dist_cwf_roy} \st{(blue points)} shows the distribution of the absorption and emission Gaussian components in their respective width and amplitude spaces. For the absorption components, a clear correlation between $\tau^p$ and $\sigma_v$ is evident. The double-humped nature of the distributions results from the two-phase nature of the medium in the simulation. The width of the emission components also shows a double-humped distribution. The $T_B$ distribution peaks at around $10\mathrm{\ K}$, which mostly arises from the warm components, and there are only a few components with $T_B<1\mathrm{\ K}$. The emission-only components turn out to be broader than the absorption components, as expected. The median values of the line width ($\sigma_v$) are $2.3$ and $5.7$ $\mathrm{km~s^{-1}}$ for the joint absorption-emission and emission-only components, respectively. 

\subsection{Using both absorption and emission components}\label{subsec:using_both_abs_em}

\begin{figure}
    \centering
    \includegraphics[width=\linewidth]{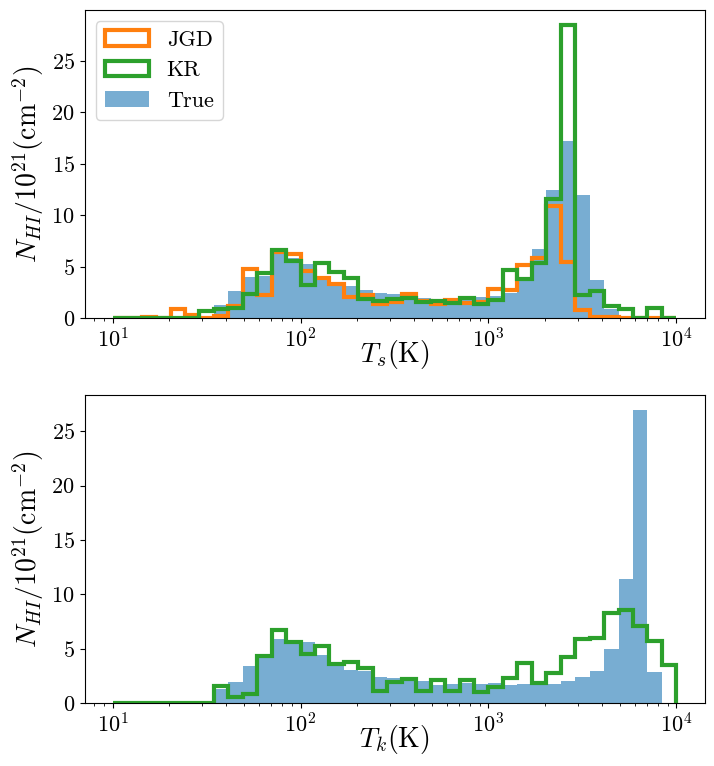}
    \caption{Inferred column density distributions (step histograms) compared to the true distribution (filled histogram). \textbf{Top: }Inferred distribution of the spin temperature using both the JGD method (\S\ref{subsec:using_both_abs_em}) and KR method on absorption components (\S\ref{subsec:using_only abs}). \textbf{Bottom: }Inferred distribution of kinetic temperatures of the absorption components using the KR method. The inferred distributions provide a reasonable match to the true distributions for temperatures \mbox{$\lesssim2500\,\mathrm{K}$}, but are comparatively less reliable at warmer temperatures ($\gtrsim2500\,\mathrm{K}$).}
    \label{fig:nh_dist_cwf_jgd_cwf}
\end{figure}

For the joint absorption-emission components, we estimate their spin temperature using the following relation:
\begin{equation}\label{eq:comp_ts_min}
    T_s^{\rm \rm comp} = \frac{T_B^p}{1-e^{-\tau^p}},
\end{equation}
and the corresponding column density as:
\begin{equation}\label{eq:comp_nh_min}
    N_{\text{H~{\sc i}}}^{\rm \rm comp}/10^{21} = 1.823\times 10^{-3}\mathrm{\ cm^{-2}\ }\sqrt{2\pi}T_s^{\rm \rm comp}\sigma_v\tau^p.
\end{equation} 
We study the inferred column density distribution over spin temperature and compare it with the simulation. The top panel of Figure \ref{fig:nh_dist_cwf_jgd_cwf} (orange step histogram) shows the JGD inferred column density distribution over $T_s^{\rm \rm comp}$. The filled histogram represents the actual distribution for the $200$ sightlines considered. We see that at cooler temperatures, $T_s\leq2500\mathrm{\ K}$, the inferred and the true distributions agree quite well. Conversely, at warmer temperatures, a consistent underestimation of column density is evident. This is mostly due to Gaussian decomposition failing to recover several low optical depth warm components in absorption (emission-only components may make up for this deficiency; see \S\ref{subsec:emonly_column_density}). As the synthetic spectra are noiseless, it is difficult to estimate a threshold of $T_s^{\rm comp}$ for absorption components to be detected. In real spectra, the noise level can be used to compute the limit of component spin temperatures to be detectable. Nevertheless, JGD could qualitatively capture the two-phase distribution of the simulation. The inferences broadly remain the same for the maximum WF effect case, with the addition of noise or increasing the emission area to reasonable values (see \S\ref{subsec:mwf_discussion}, Appendix \ref{appendix:noise_effect} and \ref{appendix:emission_beam_effect}).

Since JGD can extract the true cold gas properties of the medium, for each line of sight we estimate the cold gas column density from Gaussian decomposition as the sum of column densities of all the components with $T_s^{\rm comp}<200\mathrm{\ K}$. We found that the inferred cold gas column density for the lines of sight agrees with the true cold column densities within a factor of two for most cases. We define the inferred fraction of cold gas as the ratio of inferred cold column density and $N_{\text{H~{\sc i}}}^{\rm iso}$. This is well constrained by the relation $f_c = T_c/T_s^{\rm los}$ with the cold temperature between $T_c=30\mathrm{\ K}$ and $T_c=150\mathrm{\ K}$. This finding agrees well with that obtained with the SPONGE observations \citep[see fig.~5 in][]{Murray18}.

We test the ability of JGD to infer the gas phase (CNM, UNM, and WNM) fractions of the ISM. Table \ref{tab:gas_fractions} summarizes the inferred phase fraction with all the methods and cases. JGD cannot estimate the kinetic temperature of the components; thus, the standard definition of phases (based on $T_k$) mentioned in \S\ref{sec:simulations} cannot be used. We instead adopt the definition used in \citet{Murray18} based on the spin temperature: $T_s<250\mathrm{\ K}$ for the CNM, $250\mathrm{\ K}<T_s<1000\mathrm{\ K}$ for the UNM and $T_s>1000\mathrm{\ K}$ for the WNM. We define the inferred phase fraction based on the spin temperature as $f_X^{\rm JGD}=\sum_{T_s\in X} N_{\text{H~{\sc i}}}^{\rm comp}/\sum N_{\text{H~{\sc i}}}^{\rm comp}$, where $X$ can be CNM, UNM or WNM. For a fair comparison among the phases, we define $F_X^{\rm JGD}=f_X^{\rm JGD}/f_X^{\rm true}$, where $f_X^{\rm true}$ denotes the true phase fractions in the simulation. Considering all the cases, we find $F_{\rm CNM}^{\rm JGD}$, $F_{\rm UNM}^{\rm JGD}$ and $F_{\rm WNM}^{\rm JGD}$ take values in the range of $1.1-1.6$, $1.1-1.8$ and $0.4-0.9$. Overall, with JGD we tend to overestimate (underestimate) the cold and unstable (warm) gas fractions. 

The inferred absolute gas fractions of the CNM and UNM may suffer from several uncertainties, resulting mostly from the uncertainties in estimating the WNM gas fraction, which is subdominant in absorption. Including the emission-only components in estimation may change the inferred gas phase fractions drastically (see \S\ref{subsec:emonly_column_density}, \S\ref{appendix:noise_effect} and Table \ref{tab:gas_fractions}). In such cases, a robust quantity of interest is the ratio of the inferred amounts of CNM and UNM. We define $f^{\rm JGD}=f_{\rm CNM}^{\rm JGD}/f_{\rm UNM}^{\rm JGD}$ and the true ratio from the simulation data as $f^{\rm true}$. We find, with the definition of phases used for JGD, $f^{\rm true}\approx2.8$. For the constant WF effect case, $f^{\rm JGD}$ lies in the range of $2.5-3.6$, which agrees with true value. With the maximum WF effect case, however, $f^{\rm JGD}$ lies in the range of $3.2-4.5$, thus, overestimating the true value. 

\subsection{Using only the absorption spectra components}\label{subsec:using_only abs}

\begin{figure}
    \centering
    \includegraphics[width=\linewidth]{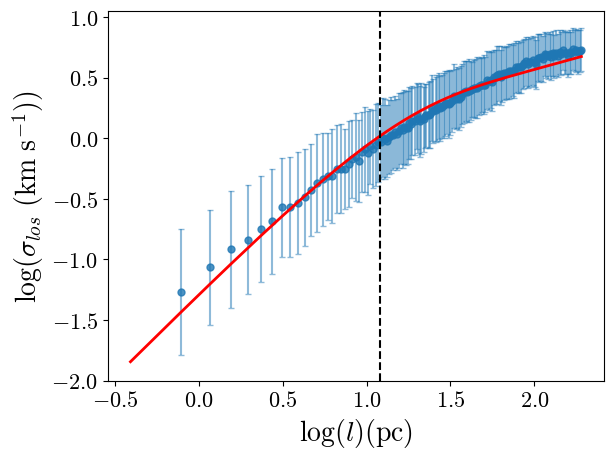}
    \caption{The true line of sight velocity dispersion in the simulation (blue points) and the model (red line) given by Equation \ref{eq:sigma_turb_sim} (with $A=0.64$, $\alpha=0.37$ and $\ell_\nu = 12\mathrm{\ pc}$). The blue points were computed by taking several 1D lines of sight inside the simulation cube of varying lengths, computing the dispersion of the $z-$axis velocity, and taking the median. The bars denote the $1\sigma$ range. The vertical dashed line corresponds to the viscous cut-off length (see Eq. \ref{eq:sigma_turb_sim}).}
    \label{fig:vel_disp_sim}
\end{figure}

Now we use the KR method, which is based on only the absorption components. It uses the statistical velocity dispersion properties of turbulence to decouple the thermal and non-thermal broadening of the spectral lines. The method estimates the column densities, spin, as well as kinetic temperatures of the absorption Gaussian components using the values of $\tau^p$ and $\sigma_v$. In their work, \citet{KR19} used an isobaric equation of state, with pressure $P$, and a Larson-like turbulent velocity scaling law \citep{Larson79}
\begin{equation}\label{eq:larson_scaling}
    \sigma_{\rm turb} = A\left(\frac{\ell}{\mathrm{pc}}\right)^{\alpha},
\end{equation}
where $A$, $\alpha$ and $P$ are parameters, whose values were taken to be $0.64\mathrm{\ km~s^{-1}}$, $0.37$ and $3700\mathrm{\ K/cm^3}$ respectively. Using these parameter values, the total column density and component spin temperatures for a selected few lines of sight (with simpler profiles: low optical depths and fewer components) more or less agree with those obtained using JGD (refer \citet{KR19} for details). 

The value of $P$ represents the average pressure of the ISM. For our work, we take $P=2717\mathrm{\ K/cm^3}$, the volume-averaged pressure of the simulation domain. A power-law turbulent velocity scaling was found to be insufficient to describe the simulation over the entire range of length scales. This is a consequence of a low Reynolds number of this simulation ($\approx2000$) compared to that estimated for the ISM in a typical galaxy \citep[$\sim10^7$, see tab.~3.1 in ][]{Brandenburg05}, which results in a large dissipation scale and an insufficient inertial range. \citet{Kriel22} found that the dissipation scale $\ell_\nu$ is related to the turbulence driving scale $\ell_{\rm turb}$ and Reynolds number $\mathrm{Re}$ as $\ell_\nu\approx40~\ell_{\rm turb}\mathrm{Re}^{-3/4}$, which in this case (with $\ell_{\rm turb}=100\mathrm{\ pc}$) is $\approx12\mathrm{\ pc}$, placing most of the simulation domain in the non-inertial turbulence regime. We find that an exponential correction term can be used at the low Reynolds numbers of our simulations
\begin{equation}\label{eq:sigma_turb_sim}
    \sigma_{\rm turb} = A\left(\frac{\ell}{\mathrm{pc}}\right)^{\alpha}\left(1-e^{-\frac{\ell}{\ell_\nu}}\right),
\end{equation}
where we have used $l_\nu = 12\mathrm{\ pc}$. Comparing Equation \ref{eq:sigma_turb_sim} with the true line of sight velocity dispersion inside the simulation domain (see Figure \ref{fig:vel_disp_sim}), shows that $A=0.64\mathrm{\ km~s^{-1}}$ and $\alpha=0.37$ (the same values used for observational data analysis) well describes the turbulence properties of the system. The value of $\alpha$ is also consistent with that observed with subsonic turbulence in simulations \citep[$0.39\pm0.02$, see][]{Federrath21}. As was done in \citet{KR19}, we checked for a few lines of sight that with this modification in $\sigma_{\rm turb}$, the total column densities and component spin temperatures match with those obtained by JGD. We note here that the modeling is fairly insensitive to the exact functional form of the scaling and the parameter values. As long as the non-thermal 1D velocity dispersion of the ISM is captured well, the inferences are unlikely to be significantly affected.  

In Figure \ref{fig:nh_dist_cwf_jgd_cwf}, we show the inferred column density distribution over spin temperature (top, green step histogram) and kinetic temperature (bottom) for the constant WF effect case. The KR method is seen to be capable of estimating the distribution quite well in the lower temperature regime ($T_s$ and $T_k$ $\lesssim 2500\mathrm{\ K}$). We also see that the inferred distribution from the KR method and JGD agree quite well in this temperature regime. At higher temperatures, the estimates are unreliable, though qualitatively, a two-phase distribution is still recovered. Here too, the inferences broadly remain the same for the maximum WF effect case or with the addition of noise (see \S\ref{subsec:mwf_discussion}, Appendix \ref{appendix:noise_effect} and \ref{appendix:emission_beam_effect}).

Similar to the JGD, we define the fractions $f_X^{\rm KR}$, $F_X^{\rm KR}$, and $f^{\rm KR}$ to study the method's ability to recover the gas phase fractions. As the KR method can give values of the kinetic temperatures of the components, we revert to the phase definition based on the kinetic temperatures as defined in \S\ref{sec:introduction}. $F_{\rm CNM}^{\rm KR}$, $F_{\rm UNM}^{\rm KR}$ and $F_{\rm WNM}^{\rm KR}$ take values in the range of $0.9-1.2$, $1.1-1.4$ and $0.6-0.9$. Thus, this method also shows the trend of overestimation of the CNM and UNM and the underestimation of WNM fractions. The value of $f^{\rm true}$, in this case, is $\approx1$. We find that $f^{\rm KR}$ lies in the range of $0.7-1.1$, which again agrees well with the true value.

\section{Discussion}\label{sec:discussions}

\subsection{Efficient Wouthuysen-Field effect}\label{subsec:mwf_discussion}
Recently, several theoretical and observational studies have indicated that the WF effect in the ISM may be efficient enough to render $T_s=T_k$ for all phases \citep{Murray18,Seon20}. To check how this might affect our inferences, we performed our analyses separately, assuming $T_s=T_k$ for all the phases (the maximum WF effect case). The results of the constant and maximum WF effect cases turned out to be broadly similar. The outcome of the data analysis techniques with the non-warm gas components remains unaltered. Both the JGD and KR methods could recover well the column densities and temperatures of gas with $T_s/T_k\lesssim2500\mathrm{\ K}$. The differences between the cases mostly lie in the warm temperature regime. Below, we discuss the key results for the case of a maximum WF effect ($T_s = T_k$) and relate them to a few observational signatures indicative of an efficient WF effect in the ISM. 

Using $T_s=T_k$ changes the distribution of spectral data in the $T_B(v)-\tau(v)$ space at the lower optical depth regime. Fitting the new distribution with our model discussed in \S\ref{subsec:tb_tau_model} yields $T_w=6000\mathrm{\ K}$ (see Eq. \ref{eq:tb_warm}), leaving the values of the other two parameters ($T_{B,w}$ and $C$) practically unchanged. This is a direct consequence of the much higher spin temperature of the warm gas in this case. The non-warm gas length scales or warm gas brightness temperature do not depend on the spin temperature, leaving them unchanged. We relate the higher value of $T_w$ in this case to that obtained by fitting the model to the observed data ($T_w=5000\mathrm{\ K}$). This is an indication of higher warm gas spin temperatures and efficient WF effect in the ISM. Higher values of parameter $T_w$, however, can also occur due to larger beam sizes associated with emission spectra than the corresponding absorption spectra (see Table \ref{tab:tb_tau_model_params} and Appendix \ref{appendix:emission_beam_effect}).

With JGD, several of the joint Gaussian components yielded very high spin temperatures ($T_s^{\rm comp}>4000\mathrm{\ K}$). This resulted in a much broader distribution of the component spin temperatures, in contrast to the constant WF effect case, where most of the spin temperatures were limited to $\approx3000\mathrm{\ K}$. Higher spin temperature results in lower optical depths of the warm components, which renders several of these components undetectable in absorption spectra. This may be a possible reason behind the non-detection of components with spin temperature in the range of $2000-3000\mathrm{\ K}$ in observations \citep{Murray15}, indicating an efficient WF effect. However, the high sensitivity absorption survey by \citet{Roy13} could detect several low-amplitude and broad components in absorption. Unfortunately, the spin temperatures of these components are not available. This, however, illustrates the importance of high-sensitivity absorption surveys to study the strength of the WF effect in the ISM.

With the maximum WF effect case, the KR method, in its original form, fails to estimate the spin temperature of the warm components, as it implicitly uses the \citet{Liszt01} relation between $T_s$ and $T_k$. Thus, for a maximum WF effect, we modify and use $T_s=T_k$ within the KR method. This improves its recovery of the true distribution for the maximum WF effect case. With this modification, however, the performance of the method worsens for a constant WF effect. This shows that the success of the method for recovering warm components with the KR method depends on our knowledge of the WF effect in the ISM. 

\subsection{Column densities of emission-only components}\label{subsec:emonly_column_density}
\begin{figure}
    \centering
    \includegraphics[width=\linewidth]{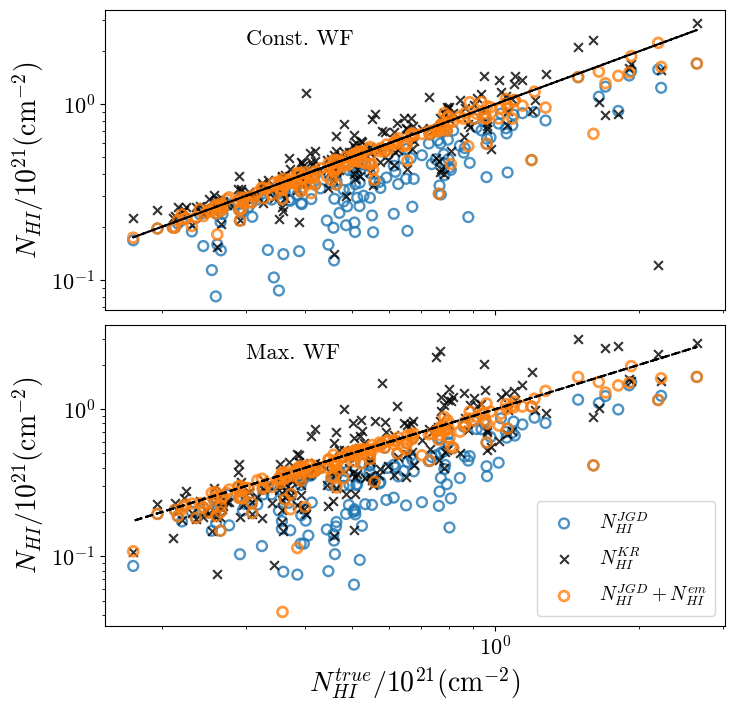}
    \caption{The inferred line of sight column density using only the joint absorption-emission components ($N_{\text{H~{\sc i}}}^{\rm JGD}$), including the emission-only components ($N_{\text{H~{\sc i}}}^{\rm JGD}+N_{\text{H~{\sc i}}}^{\rm em}$) and using only absorption components and the KR method ($N_{\text{H~{\sc i}}}^{\rm KR}$) (\textbf{Top:} constant WF effect case, \textbf{Bottom:} $T_s=T_k$ case). $N_{\text{H~{\sc i}}}^{\rm JGD}+N_{\text{H~{\sc i}}}^{\rm em}$ agrees well with the true column density. The equality lines are shown as black lines.  The KR method overestimates the line-of-sight column densities.}
    \label{fig:joint_decomp_nh_comp}
\end{figure}

Estimating the physical properties of the emission-only components is difficult due to the unavailability of the corresponding absorption components and thus, Equations \ref{eq:comp_ts_min} and \ref{eq:comp_nh_min} cannot be used. The brightness temperature of emission-only components also suffers from absorption due to the optically thicker components. Here we use a simple method to estimate the column densities of the extra components in emission spectra, assuming that they are optically thin and do not contribute to the optical depth budget. We approximate the optical depth of the optically thicker components surrounding the optically thin emission-only component to be the value of $\tau(v)$ at the velocity center of the emission-only component, which we denote as $\tau(v_{p,\rm em}^{\rm comp})$. These components are potentially responsible for the attenuation of the emission from the emission-only component. To correct for this attenuation, we assume that a fraction $F$ of the emission-only component lies in front of the optically thicker components and the rest lies behind. Using this two-phase model, we get the true peak brightness temperature estimate of the emission-only component as
\begin{equation}\label{eq:tb_true_est}
    T_B^{\rm est,peak} = \frac{T_B^p}{F+(1-F)e^{-\tau(v_{p,\rm em}^{\rm comp})}}.
\end{equation}
Statistically, we expect $F\approx 0.5$ due to the diffuse nature of the warm components and use this value in our estimation. We use $T_B^{\rm est,peak}$ and the line widths of emission-only components to estimate their column densities (using Equation \ref{eq:comp_nh_min} and using $T_s^{\rm comp}\tau^p = T_B^{\rm est,peak}$; note that here $\tau^p \leq \tau(v_{p,\rm em}^{\rm comp})$ is the optical depth corresponding to the particular emission-only component).

We denote $N_{\text{H~{\sc i}}}^{\rm em}$ as the sum of $N_{\text{H~{\sc i}}}^{\rm comp,\rm em}$ for all the emission-only components for a line of sight. Figure \ref{fig:joint_decomp_nh_comp} shows the line of sight column densities inferred from the joint absorption-emission Gaussian components only ($N_{\text{H~{\sc i}}}^{\rm JGD}$, blue points) and including the emission-only component column densities ($N_{\text{H~{\sc i}}}^{\rm JGD}+N_{\text{H~{\sc i}}}^{\rm em}$, orange points) for both the constant and maximum WF effect cases (top and bottom respectively). $N_{\text{H~{\sc i}}}^{\rm JGD}+N_{\text{H~{\sc i}}}^{\rm em}$ agrees with the true column density quite well. This serves as a verification of the completeness of the Gaussian decomposition of the spectra when emission-only components are included. This also shows that the above method gives a reasonable estimate of the emission-only components' column densities. We assume that all emission-only components comprise gas in the warm phase (in this case, $T_s>1000\mathrm{\ K}$). We find that the new inferred gas phase fractions are in excellent agreement with the true fractions (see Table \ref{tab:gas_fractions}). 

For both panels in Figure \ref{fig:joint_decomp_nh_comp}, we also include the total line of sight column density estimation from the KR method (black crosses). Compared to JGD inferred values, the KR method overestimates the column density from absorption components. This is an inherent limitation of the KR method as the parameters are chosen to approximately match the total column density in the first place (see \S\ref{subsec:using_only abs}). This leads to an overestimation of the spin temperatures, and thus column densities, of several warm components (for example, see the high inferred column density at $T_s\approx2500\mathrm{\ K}$ in Figure \ref{fig:nh_dist_cwf_jgd_cwf}), and consequently, an overestimation of the total column density.

We also perform the same analyses for the spectra with noise (see Appendix \ref{appendix:noise_effect} for the details). No significant difference is obtained in the results, except for an even more prominent underestimation of the total column density with joint absorption-emission components. However, with noise, we can get an estimate of the spin temperatures of the emission-only components. This enables us to check the changes in the inferred column density distribution when these components are included. We see that we recover a significant amount of warm gas column density. For details, see Appendix \ref{appendix:noise_effect}.

\subsection{Comparing with previous works and observations}\label{subsec:comparing_prev_works_and_obs}
\subsubsection{Properties of the synthetic Gaussian components}
We note a very good agreement between the distribution of the inferred Gaussian components in this work and a similar analysis performed by \citet{Murray17,Murray18}, using the \citet{Kim13} simulation data and the Gausspy \citep{Lindner15} decomposition algorithm \citep[see Figure \ref{fig:abs_em_comps_scatter_dist_cwf_roy} here compared to Figure 11 and 13 in][]{Murray17}. Such a similarity is surprising, given the very different simulation setups of \citet{Kim13} and this work and the different Gaussian decomposition algorithms used. This clearly shows that different ISM simulations have several underlying similarities in terms of observational signatures, which are largely successfully extracted by the Gaussian decomposition algorithms. 

Figure \ref{fig:abs_em_comps_scatter_dist_cwf_roy} shows, along with the distribution of the synthetic components, the absorption components from observations (black points from GMRT/WSRT/ATCA data; \citet{Roy13}). The latter does not follow the overall trend of the synthetic components. They also tend to populate the region in the $\tau^p-\sigma_v$ space that is only sparsely populated by the simulation components. This difference in the nature of the Gaussian components is indicative of the possible difference in the nature of the gas clouds in simulations and observations, which further illustrates the quantitative difference between the two.

\subsubsection{Temperature distribution of ISM gas}\label{subsubsec:temp_dist_ISM}

\begin{table*}
\caption{The true and inferred gas phase fractions, for various cases, along the $200$ lines of sights in the $512^3$ simulation. cWF and mWF denote the constant and maximum WF effect cases, respectively. We primarily use two definitions of gas phases. First is in terms of $T_s$ \citep[following][SPONGE21 survey]{Murray15,Murray18} and used for the JGD method. The second is in terms of $T_k$ \citep[following][GMRT/WSRT/ATCA survey]{KR19} and used for the KR method. For comparison, we provide the estimated phase fractions with these surveys. We also provide the phase fractions estimated by two other surveys: Arecibo survey \citep{Heiles03,Heiles03II,Nguyen19} and HI4PI \citep{Kalberla18}. We note that these works use different definitions of gas phase temperatures. We explicitly mention the definitions of phases used to compute the gas fraction. Additionally, we also provide the number of Gaussian components used for analysis for each of the cases.}
\label{tab:gas_fractions}
\begin{threeparttable}
\sisetup{
                detect-mode,
                tight-spacing            = true,
                group-digits             = false,
                input-signs              = ,
                input-symbols            = ,
                input-open-uncertainty   = ,
                input-close-uncertainty  = ,
                table-align-text-pre     = false,
                round-mode               = figures,
                round-precision          = 0,
                %       round-integer-to-decimal = true,
                table-space-text-pre     = (,
                table-space-text-post    = ),
            }
\centering
\resizebox{1.0\linewidth}{!}{%
\begin{tabular}{|c|c|c|c|c|c|c||c|c|c|c|} 
\hline
                                          & \multicolumn{2}{c|}{$f_\mathrm{CNM}$}        & \multicolumn{2}{c|}{$f_\mathrm{UNM}$}        & \multicolumn{2}{c||}{$f_\mathrm{WNM}$}        & \multicolumn{4}{c|}{Gaussian Components}                                               \\ 
\cline{2-11}
True: by Ts$^\text{a}$ (by Tk$^\text{b}$) & \multicolumn{2}{c|}{0.36 (0.32)} & \multicolumn{2}{c|}{0.13 (0.34)} & \multicolumn{2}{c||}{0.51 (0.34)} & \multicolumn{2}{c|}{Absorption}           & \multicolumn{2}{c|}{Emission-only}         \\ 
\hline
                                          & cWF         & mWF                & cWF         & mWF                & cWF         & mWF                & cWF       & mWF                           & cWF & mWF                                  \\ 
\cline{2-11}
JGD$^\text{a}$ (KR$^\text{b}$)            & 0.45 (0.31) & 0.48 (0.32)        & 0.15 (0.45) & 0.14 (0.41)        & 0.40 (0.24) & 0.38 (0.26)        & 532 (604) & 486 (584)                     & 263 & 288                                  \\
With Noise$^\text{e}$: JGD (KR)           & 0.61 (0.41) & 0.62 (0.38)        & 0.17 (0.42) & 0.16 (0.35)        & 0.23 (0.17) & 0.22 (0.27)        & 375 (425) & 340 (417)                     & 383 & 396                                  \\
With N=3$^\text{f}$: JGD                  & 0.42        & 0.46               & 0.15        & 0.13               & 0.43        & 0.41               & 501       & 486                           & 276 & 257                                  \\
With N=5$^\text{f}$: JGD                  & 0.40        & 0.45               & 0.13        & 0.10               & 0.46        & 0.45               & 488       & 428                           & 271 & 292                                  \\
With N=15$^\text{f}$: JGD                  & 0.28        & 0.33               & 0.08        & 0.08               & 0.64        & 0.58               & 465       & 404                           & 274 & 292                                  \\
With Emission-only$^\text{g, j}$: JGD        & 0.35        & 0.35               & 0.12        & 0.10               & 0.53        & 0.55               & --        & --                            & --  & --                                   \\ 
\hline
GMRT/WSRT/ATCA (30 LOS)$^\text{b, h}$      & \multicolumn{2}{c|}{0.15}        & \multicolumn{2}{c|}{0.75}        & \multicolumn{2}{c||}{0.1}         & \multicolumn{2}{c|}{214}                  & \multicolumn{2}{c|}{–}                     \\
SPONGE (57 LOS)$^\text{a, i}$              & \multicolumn{2}{c|}{0.56}        & \multicolumn{2}{c|}{0.41}        & \multicolumn{2}{c||}{0.03}        & \multicolumn{2}{c|}{\multirow{2}{*}{280}} & \multicolumn{2}{c|}{\multirow{2}{*}{278}}  \\
SPONGE+Emission-only$^\text{a, j}$        & \multicolumn{2}{c|}{0.28}        & \multicolumn{2}{c|}{0.20}        & \multicolumn{2}{c||}{0.51}        & \multicolumn{2}{c|}{}                     & \multicolumn{2}{c|}{}                      \\ 
\hline
Arecibo (77 LOS)~$^\text{c, j}$~          & \multicolumn{2}{c|}{0.4}         & \multicolumn{2}{c|}{0.24}        & \multicolumn{2}{c||}{0.36}        & \multicolumn{2}{c|}{349}                  & \multicolumn{2}{c|}{327}                   \\
HI4PI~$^\text{d, k}$                   & \multicolumn{2}{c|}{0.24}           & \multicolumn{2}{c|}{0.44}        & \multicolumn{2}{c||}{0.34}           & \multicolumn{2}{c|}{–}                    & \multicolumn{2}{c|}{--}                    \\
\hline
\end{tabular}
}
\begin{tablenotes}
       \item [a] CNM: $T_s<250\mathrm{\ K}$; UNM: $250\mathrm{\ K}<T_s<1000\mathrm{\ K}$; WNM: $T_s>1000\mathrm{\ K}$
       \item [b] CNM: $T_k<200\mathrm{\ K}$; UNM: $200\mathrm{\ K}<T_k<5000\mathrm{\ K}$; WNM: $T_k>5000\mathrm{\ K}$
       \item [c] CNM: $T_{\rm max}<500\mathrm{\ K}$; UNM: $500\mathrm{\ K}<T_{\rm max}<5000\mathrm{\ K}$; WNM: $T_{\rm max}>5000\mathrm{\ K}$ where $T_{\rm max}=121\sigma_v^2$, these lines of sights are closer to giant molecular clouds
       \item [d] CNM: $T_{\rm max}=283\pm15\mathrm{\ K}$, UNM: $T_{\rm max}=2014\pm543\mathrm{\ K}$, WNM: $T_{\rm max}=11835\pm598\mathrm{\ K}$
       \item [e] See Appendix \ref{appendix:noise_effect}
       \item [f] See Appendix \ref{appendix:emission_beam_effect}
       \item [g] See \S\ref{subsec:emonly_column_density}
       \item [h] Uses the absorption components only
       \item [i] Uses the joint absorption-emission components
       \item [j] Uses both the joint absorption-emission and emission-only components
       \item [k] Uses only emission components
     \end{tablenotes}
  \end{threeparttable}
\end{table*}

In this section, we compare the amount of gas in different phases as inferred by various observational surveys from 21~cm spectral data and as produced in the current simulations. Table~\ref{tab:gas_fractions} lists the column density fraction (equivalent to the mass fraction) of CNM, UNM, and WNM inferred by four surveys: the GMRT/WSRT/ATCA \citep{KR19}, SPONGE21 \citep{Murray18}, Arecibo \citep{Heiles03II, Nguyen19} and HI4PI \citep{Kalberla18}, along with the number of sightlines used and the number of extracted Gaussian components, wherever applicable. It is seen readily that the inferred phase fractions across different surveys show significant differences (for example, the estimated $f_{\rm UNM}$ from GMRT/WSRT/ATCA vs. SPONGE21 in Table~\ref{tab:gas_fractions}), which suggests that the results from different surveys are inconsistent. However, it is important to note that the gas phase definitions (temperature boundaries and whether spin or kinetic temperature) used in the various surveys often differ significantly (see footnotes in the table). The gas phase fractions can be quite different with different phase definitions (for example, see the phase fractions in the simulation domain with two different phase definitions, first row in Table \ref{tab:gas_fractions}). Thus, it is difficult to compare the inferred phase fractions across different surveys. This establishes the importance of using a common (or at least similar) definition of the gas phases across different studies (both simulations and observations). A similar need was also noted in \citet{Griffiths23}.

There are, however, a few factors that may make the observationally inferred phase fractions (or, in general, the temperature distribution) across surveys different to some degree. Firstly, the inferred cold and unstable gas fractions depend heavily on how the warm gas fraction is estimated. If the emission-only components are used to estimate the warm gas missing in absorption (similar to as discussed in \S\ref{subsec:emonly_column_density}), the inferred phase fractions change significantly (see SPONGE vs SPONGE+Emission-only in Table \ref{tab:gas_fractions}). Other than this, the analysis model (see \S\ref{subsubsec:lim_of_obs_analy_techs} for a discussion) or the choice of line of sight (for example, low vs.~high galactic latitude or proximity to molecular clouds) may also result in differently inferred phase fractions.

Despite the uncertainties involved with observational estimates, we attempt to bring out the commonalities among the results from different surveys and compare the observationally inferred gas-phase distribution with that in the simulation. For comparison, we choose the GMRT/WSRT/ATCA and SPONGE21 survey results for primarily two reasons: I) their analysis methods are similar to this work, and II) they sample the ISM across all galactic latitudes roughly uniformly, and thus do not suffer from the bias arising from the specific choice of lines of sight. Though we list the inferred phase fractions by Arecibo and HI4PI in the table for completeness, henceforth, we do not consider them owing to their very different analysis methods and/or gas phase definitions. On the other hand, results from a variety of ISM simulations, including large-scale ISM simulations, are available in the literature. Quantitatively, the different simulations often show different results. Here, too, we try to refer to the common features in all the simulations whenever possible.

Comparison of the gas properties in the simulation with those inferred from observations reveals a few quantitative differences, mostly related to the amount of cold and unstable gas. We first consider the $f_{\rm CNM}$ and $f_{\rm UNM}$ inferred from observations. As already discussed briefly, these values are significantly affected by the ways the warm gas fraction is estimated. To avoid this uncertainty, we instead consider the ratio of cold and unstable gas (i.e., $f_{\rm CNM}/f_{\rm UNM}$). Firstly, we note that \citet{KR19} estimates a significantly high fraction of unstable gas ($\sim75\%$, $200\mathrm{\ K}<T_k<5000\mathrm{\ K}$) and a very low amount of cold gas ($\sim15\%$, $T_k<200\mathrm{\ K}$). The ratio of the amount of cold and unstable gas, thus, is $\approx0.2$. This is significantly lower than that for the simulation ($\approx1$ with the phase definitions with $T_k$, see \S\ref{subsec:using_only abs}). Similarly, with JGD, \citet{Murray18} estimates unstable ($250\mathrm{\ K}<T_s<1000\mathrm{\ K}$) and cold ($T_s<250\mathrm{\ K}$) gas fractions of $41\%$ ($20\%$) and $56\%$ ($28\%$), respectively, without (with) including the emission-only components. Here, the ratio of the amount of cold and unstable gas, thus, is $\approx1.4$. This is again significantly lower than that of the simulation when the phase definition with $T_s$ is used ($\approx2.8$, see \S\ref{subsec:using_both_abs_em}).

The quantity $f_{\rm CNM}/f_{\rm UNM}$ is still dependent on the definition of gas phases. Thus, we now check the observationally inferred column density distribution itself. Although there is a significant amount of unstable gas in the simulation, the lower temperature peak in the distribution occurs at $T_k \text{ or } T_s\approx100\mathrm{\ K}$, which lies in the temperature range of CNM. The distribution is also two-phase-like, with distinct peaks in the CNM and WNM regimes. Several large-scale ISM simulations also show similar behavior \citep{Saury14,Kim13,Kim17,Rathjen21,Kim23}. We contrast this with the distributions obtained from observations in \citet[][Figure 8]{Murray18} and \citet[][Figure 6]{KR19}. Though the two distributions may differ quantitatively, they share the common feature of lacking the prominent two-peak feature. In fact, both the inferred distributions tend to peak in the UNM temperature regime, contrary to almost all simulations.

We now use the results of this work to show that the above differences are unlikely to arise solely from the uncertainties related to observational data analysis and the inferences thereof. Our analysis shows that both JGD and KR methods (even when the various observational effects are considered, see Appendix \ref{appendix:noise_effect}, \ref{appendix:emission_beam_effect} and Table \ref{tab:gas_fractions}) are reliable in recovering the gas distribution of these phases reasonably well in most cases (see Figure \ref{fig:nh_dist_cwf_jgd_cwf}, and also Figures \ref{fig:nh_dist_cwf_jgd_cwf_with_noise}, \ref{fig:jgd_3c3_5c5}). Only with a very large emission beam averaging (over a rectangular beam of $\sim~6~\mathrm{\ pc}$, see Appendix \ref{appendix:emission_beam_effect}) JGD fails to recover the true distribution well. But such a case is unphysical, as argued in Appendix \ref{appendix:emission_beam_effect}, and unlikely to occur with observations. Even then, a qualitative two-phase distribution is recovered, and the inferred gas phase fractions are broadly consistent with the other cases (see Table \ref{tab:gas_fractions}). The analysis methods, in any case, do not tend to overestimate the amount of unstable gas over the cold gas (inferred $f_{\rm CNM}/f_{\rm UNM}$ is always $\geq$ the true value, see the values of $f^{\rm JGD}$ and $f^{\rm KR}$ in \S\ref{subsec:using_both_abs_em} and \S\ref{subsec:using_only abs} respectively). Thus, at this stage, the indication of a higher (lower) amount of unstable (cold) gas in ISM from observations seems improbable to be an artifact of the analysis methods. 

There is also an additional discrepancy with the WNM gas. The inferred warm gas fraction in observations using both JGD (without including emission-only components) and KR methods is surprisingly low (see $f_{\rm WNM}$ for ``SPONGE" and ``GMRT/WSRT/ATCA" rows in Table \ref{tab:gas_fractions}). Although our study suggests that we tend to underestimate the warm gas fraction with these analyses (e.g., some emission-only components are missing in absorption), it still does not explain the very low observationally inferred warm gas fractions. Observations suggest that warm gas is unexpectedly subdominant in absorption spectra. This may occur if the warm gas has very high temperatures through efficient WF effect, which decreases its optical depth. However, in our study, even with the maximum WF effect case, we do not see such a severe underestimation of warm gas. This puzzle was also noted by \citet{Murray18}. This behavior is clearly still an open question and is another point of difference between the simulations and observations. 

When the emission-only components are included in JGD analysis, the inferred phase fractions change significantly (see ``SPONGE" and ''SPONGE+Emission-only" rows in Table \ref{tab:gas_fractions}). Our study shows that inclusion of emission-only components can help us recover the column density of warm gas, which is otherwise subdominant in absorption spectra (see Table \ref{tab:gas_fractions} and Figures \ref{fig:joint_decomp_nh_comp} and \ref{fig:nh_dist_cwf_jgd_cwf_with_noise_with_emonly_0.002}). In fact, the inferred warm gas fraction by SPONGE21 with emission-only components is in very good agreement with that in the simulation (see Table \ref{tab:gas_fractions}), and the values of $f_{\rm CNM}$ and $f_{\rm UNM}$ also come closer to that in the simulation.   

\subsection{Limitations and Future Scopes}\label{subsec:mismatch}

The differences between the simulations and observational inferences, as discussed in the preceding section, are mostly open questions. The limitations of the current simulations, both in terms of capturing all the essential ISM physics or insufficient numerical resolution, may lead to such disagreements. There may be several shortcomings with observations, too, which may lead to erroneous inferences and are yet to be properly studied. Below, we discuss some of the aspects that demand future work in order to better understand and/or resolve these discrepancies.

\subsubsection{Simulations}\label{subsubsec:lim_of_sims}
There are mainly two sources of limitations in simulations - the ability to include all essential ISM physics and the effect of numerical resolution. We discuss them separately.

\underline{\emph{ISM Physics:}} Several small-scale simulations aimed at studying ISM properties, like the one used in this work, potentially suffer from inaccuracies with regard to the proper implementation of the various physical processes occurring in ISM, especially on larger scales. One such aspect is the way turbulence is driven in the medium. These simulations use a constant turbulent forcing uniformly for the whole simulation domain. However, this may not be %a proper
an appropriate description of turbulence driving in ISM, where transient supernova explosions are one of the main sources of turbulence. The gas temperature distribution depends on the position of the supernovae relative to the ISM thermal phases \citep{Gatto15}. Supernovae exploding within cold clouds may heat the surrounding gas, leading to higher amounts of unstable phase gas. The intermittent nature of supernovae may also influence the amounts of cold and unstable gas: a significant amount of unstable gas may form in the decaying phase of the turbulence when the driving is not active \citep[see, for example, Figure 12 in][]{Saury14}. Additionally, strong magnetic fields generated at supernova shock fronts may stabilize thermal instability (e.g., \citealt{Sharma2010}) preventing the formation of cold clouds and leading to higher amounts of isobarically unstable phase gas \citep[for observational evidence of high magnetic field regions see][]{Thomson19,Bracco20}. Other than this, underestimation of the strength of turbulence forcing in the ISM can affect the gas distribution to a large extent. Increasing the turbulence strength can increase (decrease) the amount of unstable (cold) gas and can also lead to a complete wipe-out of the bimodal gas distribution \citep{Seifried11}. Galactic potential also affects the gas properties in ISM. It leads to the accumulation of denser cold clouds nearer to the galactic plane, which may lead to higher amounts of unstable or warm gas at higher latitudes. Such effects are not captured in small-scale simulations due to the absence of an applied galactic potential.

\underline{\emph{Numerical Resolution:}} Insufficient simulation resolution often does not lead to %proper 
convergence of the amount of gas in different phases and may lead to incorrect conclusions. We refer to the resolution study in this work (Appendix \ref{app:resolution_effect}), which shows that with finer resolution, the amount of cold (unstable) gas increases (decreases) significantly when the other simulation conditions are kept unchanged. Though this shows that increasing resolution may not solve the problem of lower unstable gas in simulations (rather make it worse), it is necessary to attain convergence for comparison with other studies and observations. The large-scale ISM simulations, like those by \citet[][TIGRESS-NCR and previous works]{Kim23} or \citet[][SILCC VI and previous works]{Rathjen21}, which otherwise incorporate more known ISM physics than us, suffer from this limitation. With resolutions of the order of a few parsecs, the cold and unstable gas regime is often not well-resolved, which may potentially lead to inaccurate cold and unstable gas fractions in such simulations.

We now note that the two sources of limitations discussed above cannot be resolved individually. To see how large-scale ISM physics affects the temperature distribution of the gas, it is essential to devise ways to incorporate such effects within high-resolution small-scale simulation domains, which can resolve all the essential gas structures. In other words, it is necessary to bridge the gap between the several large and small-scale simulations. Such a setup will help us to quantify the effect of various ISM physics on the amount of gas in different phases better and %better 
to understand the origin of the disagreements between simulations and observations.

\subsubsection{Observations}\label{subsubsec:lim_of_obs_analy_techs}

Below, we discuss three main limitations associated with the observations. We note that our study includes the effects of the first two, and we do not witness significant alteration of the inferences in realistic scenarios. However, we acknowledge the possibility of larger effects in observation, which require further tests with different types of simulations. 

\underline{\emph{Larger Emission Beam Width:}} One of the main limitations associated with the observations is the larger emission beam widths. Being mostly small-scaled, cold gas can be affected by averaging over larger areas as it can reduce their significance in the spectra, potentially leading to their underestimation (see Appendix \ref{appendix:emission_beam_effect}). Our study with a very large assumed emission beam width (averaging the emission spectra over $15\times15$ pixels around the single-pixel absorption spectra) shows that such effects can potentially lead to significant errors in the estimation of the column density distribution of the gas. Though such extreme beam smearing is unlikely to occur in observations (with the spatial resolution achieved with the current surveys, see Appendix \ref{appendix:emission_beam_effect}), it is important to recognize the errors that %they 
it may cause. We also note that, in reality, the averaging happens over a cone (corresponding to the angular beam width, unlike the implementation in this study), which may affect the inferences differently. Efforts have to be made to acquire emission spectra with beam sizes as small as possible and quantify the corrections from this effect through analysis with simulations. A comparison of observations with different emission beam sizes will also be useful to quantify the uncertainties.

\underline{\emph{Spectral Blending of Incoherent Components:}} The data analysis techniques involving Gaussian decomposition have several limitations, though they are able to recover the ISM properties broadly. One of them is velocity crowding, where the spectral profiles of spatially incoherent components blend together to give a single apparent Gaussian component, leading to erroneous inferences \citep{Dickey90}. Several attempts have been made to correct this in spectral line fitting \citep{Haud00,Martin15,MD17,Marchal19}. Another similar effect has also been noted in several studies where spatially incoherent structures at a velocity channel can blend to give apparent high-density structures, which may not arise from true cold components \citep{Lazarian20,Fukui18,Hu23}. We encountered yet another effect in our work. We found that the blending may also take place between spatially coherent structures but with very different temperatures (thermally incoherent; in some cases, the blended components' temperatures were seen to span an order of magnitude). This breaks down the assumption that the Gaussian components are isothermal. The inferred spin temperature can be close to the harmonic mean of all the blended components, hence biased towards the low-temperature values. Such erroneous temperature estimations may affect the recovery of warm gas properties from the spectra.  The frequency of such blending or its effect on the inferred properties, however, remains to be quantified.

\underline{\emph{Analysis Model Dependence:}} The dependence of the quantitative inferences on the adopted analysis methodology also cannot be ruled out, especially those involving Gaussian decomposition, which may suffer from having degenerate solutions. Additionally, the success of these methods may vary across systems with different gas morphology. Such a possibility has already been noted in \citet{Murray17}. Using hydrodynamic simulations with a galactic potential by \citet{Kim13}, they showed that their JGD method does not work equally well for different latitudes \citep[see \S5.3 in][]{Murray17}. On this front, the adoption of a common set of data analysis methodologies across different studies is preferable, and they should be tested against different simulations.

\section{Conclusions}\label{sec:conclusion}
Several analysis methods are regularly applied on ISM \text{H~{\sc i}} 21cm observational data to infer the temperature and column density of the gas. These methods, however, need verification against simulation data and synthetic spectra to ascertain their reliability. In this work, we consider analysis methods both with and without the Gaussian decomposition of the spectra and test them against synthetic spectra generated from MHD simulations by \citet{Seta22}. Our work also resulted in new physical models to explain the behaviour of the spectral data. Recognizing the uncertainty regarding the extent of the Wouthuysen-Field (WF) effect in the ISM, we have tested the analysis methods and models with two choices of the WF effect: constant \citep[following][]{Liszt01} and maximum ($T_s=T_k$). We primarily perform our analysis with noiseless data. However, to test how practical problems associated with observations may affect the inferences, we consider two other cases: spectra with noise (Appendix \ref{appendix:noise_effect}) and spectra with larger emission beam width than absorption (Appendix \ref{appendix:emission_beam_effect}). To test the stability of our methods and the convergence of the results, we also consider simulations with different resolutions (Appendix \ref{app:resolution_effect}). Below, we summarize our main results:

\begin{itemize}

\item We develop a new quantitative model that explains the right boundary of the $T_B(v)-\tau(v)$ distribution and fits both the simulation and observed ISM data quite well (see Figure~\ref{fig:tb_tau_inner_both}). The model yields a relation between the maximum non-warm (cold and unstable) gas temperature, $T_{nw}$ and the optical depth $\tau$: $T_{nw}\sim\tau^{-0.4}$, which is similar to classical $T_s-\tau$ correlations. On the other hand, a single model cannot explain the left boundary of the distribution for both the simulation and observed data. This is possibly due to the different cold gas morphology between the true and simulated ISM. Further extensions of our model and detailed comparison between ISM and synthetic data may constrain the ISM properties, including the nature of turbulence as well as the presence of very small-scale or high-velocity cold gas.

\item We perform joint Gaussian decomposition (JGD) of the synthetic absorption and emission spectral lines. The properties of the inferred Gaussian components (Figure~\ref{fig:abs_em_comps_scatter_dist_cwf_roy}) are qualitatively similar to those obtained by earlier work, despite the different simulation conditions and analysis algorithms. This demonstrates the underlying commonality between different ISM simulations. 
JGD could recover the column density distribution of gas with $T_s\lesssim2500\mathrm{\ K}$ quite well, whereas the warm column densities are underestimated (Figure~\ref{fig:nh_dist_cwf_jgd_cwf}). We further show that the inclusion of emission-only components (components recovered from emission spectra but undetectable in absorption) in column density estimation can help recover most of the warm gas column density (Figure \ref{fig:joint_decomp_nh_comp} and also Figure \ref{fig:nh_dist_cwf_jgd_cwf_with_noise_with_emonly_0.002} for spectra with noise).

\item We test the newly proposed method by \citet{KR19} to extract the kinetic temperature and column densities of the Gaussian components using only the absorption spectra. Like the JGD method, this method too can recover well the properties (temperature and column density) of gas with $T_s/T_k\lesssim2500\mathrm{\ K}$ (Figure \ref{fig:nh_dist_cwf_jgd_cwf}). At higher temperatures, the estimations become unreliable. The success of this method with the warm components is dependent on the knowledge of the strength of the WF effect (see discussion in \S\ref{subsec:mwf_discussion}). 

\item With noise and larger emission beam widths, though broadly, our results and inferences remain unchanged, the inferences suffered from quantitative differences (see Appendix \ref{appendix:noise_effect} and \ref{appendix:emission_beam_effect} and the related figures). We show that larger emission beam widths may alter the parameter values of our model for the $T_B(v)-\tau(v)$ distribution and may lead to less accurate estimation of cold gas column densities with the JGD method. On the other hand, noise adds to the underestimation of warm gas by further suppressing the low-amplitude components in absorption.

\item In observations, the inferred gas phase fractions from different surveys are often not consistent. The different gas phase definitions used in the different surveys may significantly contribute to such apparent inconsistencies. This establishes the need to adopt a common set of gas-phase definitions. Other factors like the analysis model dependence of the inference or bias from the choice of line of sights may contribute to quantitative differences in both phase fractions and temperature distribution. These factors demand corrections. Despite these, inferences from several surveys show common features that we compare to the simulation properties.  

Comparison of the inferences from SPONGE21 \citep[][which uses the JGD method]{Murray18} and GMRT/WSRT/ATCA survey \citep[][which uses the KR method]{KR19} with the simulation shows quantitative disagreements (see \S\ref{subsubsec:temp_dist_ISM} for the detailed discussion). The JGD-inferred column density distributions from both surveys lack the two-phase property, as seen in this and several other simulations, and tend to peak in the UNM regime. We also show that the ratio of the amount of CNM to UNM inferred from observations is a factor of $\sim 2-5$ (depending on the gas phase definition) lower than in the simulation. Thus, observations suggest higher (lower) amounts of unstable (cold) gas than in the simulation. Our analysis shows that these disagreements are unlikely to be just an artifact of the observational data analysis methods.

\item We discuss the scope of future studies to understand better the origin of the above-mentioned discrepancy between simulations and observations. We establish the need to develop ways to incorporate the large-scale turbulence driving effects in ISM (for example, intermittent supernova explosions) in smaller high-resolution simulation domains with resolution convergence studies as done in this work (see Appendix \ref{app:resolution_effect}). With observations, the analysis model dependence of the inferences needs to be quantified through verification against simulations. Such efforts will enable checking for consistency of inferences across various surveys and more robust comparisons of simulations and observations.

\end{itemize}

\section*{Acknowledgements}
We thank the anonymous referee for their constructive comments. S.~B.~acknowledges the support from the Kishore Vaigyanik Protsahan Yojana (KVPY) scheme of the Department of Science and Technology, Government of India, a former fellowship program for undergraduate studies in basic science. C.~F.~acknowledges funding provided by the Australian Research Council (Future Fellowship FT180100495 and Discovery Project DP230102280), and the Australia-Germany Joint Research Cooperation Scheme (UA-DAAD). We further acknowledge high-performance computing resources provided by the Leibniz Rechenzentrum and the Gauss Centre for Supercomputing (grants~pr32lo, pr48pi and GCS Large-scale project~10391), the Australian National Computational Infrastructure (grant~ek9) and the Pawsey Supercomputing Centre (project~pawsey0810) in the framework of the National Computational Merit Allocation Scheme and the ANU Merit Allocation Scheme.

We are grateful to Harvey Liszt for providing N.R. with his simulation results. We have used softwares Numpy \citep{harris2020array}, SciPy \citep{2020SciPy-NMeth}, Matplotlib \citep{Hunter:2007}, and Pandas \citep{reback2020pandas} at various stages of this research. This research has made use of NASA’s Astrophysics
Data System. We acknowledge the use of data products from the Leiden/Argentina/Bonn Galactic H~{\sc i} survey and the ATCA/GMRT/WSRT H~{\sc i} absorption survey for this work. We are also grateful to the anonymous reviewer for carefully reviewing our manuscript and providing constructive comments.

%%%%%%%%%%%%%%%%%%%%%%%%%%%%%%%%%%%%%%%%%%%%%%%%%%
\section*{Data Availability}

The 21 cm data used for this study are available publicly from the ATCA, GMRT and WSRT archives and the AIfA H~{\sc i} Surveys Data Server. 
The simulation data will be provided upon reasonable request to the corresponding author, AS, or CF. All the derived quantities and models produced in this study will be shared on reasonable request to the corresponding author. 



% if your bibtex file is called example.bib

%\begin{thebibliography}{99}
%\end{thebibliography}
\bibliographystyle{mnras}
\bibliography{example} 

\begin{thebibliography}{}
\makeatletter
\relax
\def\mn@urlcharsother{\let\do\@makeother \do\$\do\&\do\#\do\^\do\_\do\%\do\~}
\def\mn@doi{\begingroup\mn@urlcharsother \@ifnextchar [ {\mn@doi@}
  {\mn@doi@[]}}
\def\mn@doi@[#1]#2{\def\@tempa{#1}\ifx\@tempa\@empty \href
  {http://dx.doi.org/#2} {doi:#2}\else \href {http://dx.doi.org/#2} {#1}\fi
  \endgroup}
\def\mn@eprint#1#2{\mn@eprint@#1:#2::\@nil}
\def\mn@eprint@arXiv#1{\href {http://arxiv.org/abs/#1} {{\tt arXiv:#1}}}
\def\mn@eprint@dblp#1{\href {http://dblp.uni-trier.de/rec/bibtex/#1.xml}
  {dblp:#1}}
\def\mn@eprint@#1:#2:#3:#4\@nil{\def\@tempa {#1}\def\@tempb {#2}\def\@tempc
  {#3}\ifx \@tempc \@empty \let \@tempc \@tempb \let \@tempb \@tempa \fi \ifx
  \@tempb \@empty \def\@tempb {arXiv}\fi \@ifundefined
  {mn@eprint@\@tempb}{\@tempb:\@tempc}{\expandafter \expandafter \csname
  mn@eprint@\@tempb\endcsname \expandafter{\@tempc}}}

\bibitem[\protect\citeauthoryear{{Audit} \& {Hennebelle}}{{Audit} \&
  {Hennebelle}}{2005}]{Audit05}
{Audit} E.,  {Hennebelle} P.,  2005, \mn@doi [\aap]
  {10.1051/0004-6361:20041474}, \href
  {https://ui.adsabs.harvard.edu/abs/2005A&A...433....1A} {433, 1}

\bibitem[\protect\citeauthoryear{{Basu}, {Roy}, {Beuther}, {Syed}, {Ott},
  {Soler}, {Stil}  \& {Rugel}}{{Basu} et~al.}{2022}]{Basu22}
{Basu} A.,  {Roy} N.,  {Beuther} H.,  {Syed} J.,  {Ott} J.,  {Soler} J.~D.,
  {Stil} J.,   {Rugel} M.~R.,  2022, \mn@doi [\mnras] {10.1093/mnras/stac3043},
  \href {https://ui.adsabs.harvard.edu/abs/2022MNRAS.517.5063B} {517, 5063}

\bibitem[\protect\citeauthoryear{{Beattie}, {Krumholz}, {Skalidis},
  {Federrath}, {Seta}, {Crocker}, {Mocz}  \& {Kriel}}{{Beattie}
  et~al.}{2022}]{Beattie22}
{Beattie} J.~R.,  {Krumholz} M.~R.,  {Skalidis} R.,  {Federrath} C.,  {Seta}
  A.,  {Crocker} R.~M.,  {Mocz} P.,   {Kriel} N.,  2022, \mn@doi [\mnras]
  {10.1093/mnras/stac2099}, \href
  {https://ui.adsabs.harvard.edu/abs/2022MNRAS.515.5267B} {515, 5267}

\bibitem[\protect\citeauthoryear{{Bracco}, {Jeli{\'c}}, {Marchal}, {Turi{\'c}},
  {Erceg}, {Miville-Desch{\^e}nes}  \& {Bellomi}}{{Bracco}
  et~al.}{2020}]{Bracco20}
{Bracco} A.,  {Jeli{\'c}} V.,  {Marchal} A.,  {Turi{\'c}} L.,  {Erceg} A.,
  {Miville-Desch{\^e}nes} M.~A.,   {Bellomi} E.,  2020, \mn@doi [\aap]
  {10.1051/0004-6361/202039283}, \href
  {https://ui.adsabs.harvard.edu/abs/2020A&A...644L...3B} {644, L3}

\bibitem[\protect\citeauthoryear{{Brandenburg} \& {Subramanian}}{{Brandenburg}
  \& {Subramanian}}{2005}]{Brandenburg05}
{Brandenburg} A.,  {Subramanian} K.,  2005, \mn@doi [\physrep]
  {10.1016/j.physrep.2005.06.005}, \href
  {https://ui.adsabs.harvard.edu/abs/2005PhR...417....1B} {417, 1}

\bibitem[\protect\citeauthoryear{{Chengalur}, {Kanekar}  \& {Roy}}{{Chengalur}
  et~al.}{2013}]{Chengalur13}
{Chengalur} J.~N.,  {Kanekar} N.,   {Roy} N.,  2013, \mn@doi [\mnras]
  {10.1093/mnras/stt658}, \href
  {https://ui.adsabs.harvard.edu/abs/2013MNRAS.432.3074C} {432, 3074}

\bibitem[\protect\citeauthoryear{{Dickey} \& {Benson}}{{Dickey} \&
  {Benson}}{1982}]{Dickey82}
{Dickey} J.~M.,  {Benson} J.~M.,  1982, \mn@doi [\aj] {10.1086/113103}, \href
  {https://ui.adsabs.harvard.edu/abs/1982AJ.....87..278D} {87, 278}

\bibitem[\protect\citeauthoryear{{Dickey} \& {Lockman}}{{Dickey} \&
  {Lockman}}{1990}]{Dickey90}
{Dickey} J.~M.,  {Lockman} F.~J.,  1990, \mn@doi [\araa]
  {10.1146/annurev.aa.28.090190.001243}, \href
  {https://ui.adsabs.harvard.edu/abs/1990ARA&A..28..215D} {28, 215}

\bibitem[\protect\citeauthoryear{{Dickey}, {Strasser}, {Gaensler}, {Haverkorn},
  {Kavars}, {McClure-Griffiths}, {Stil}  \& {Taylor}}{{Dickey}
  et~al.}{2009}]{Dickey09}
{Dickey} J.~M.,  {Strasser} S.,  {Gaensler} B.~M.,  {Haverkorn} M.,  {Kavars}
  D.,  {McClure-Griffiths} N.~M.,  {Stil} J.,   {Taylor} A.~R.,  2009, \mn@doi
  [\apj] {10.1088/0004-637X/693/2/1250}, \href
  {https://ui.adsabs.harvard.edu/abs/2009ApJ...693.1250D} {693, 1250}

\bibitem[\protect\citeauthoryear{{Draine}}{{Draine}}{2011}]{Draine11}
{Draine} B.~T.,  2011, {Physics of the Interstellar and Intergalactic Medium}

\bibitem[\protect\citeauthoryear{Elise~Albert \& Danly}{Elise~Albert \&
  Danly}{2005}]{EliseAlbert2005}
Elise~Albert C.,  Danly L.,  2005, Interemdiate-velocity Clouds.
Springer Netherlands, Dordrecht, pp 73--100, \mn@doi{10.1007/1-4020-2579-3_4},
  \url {https://doi.org/10.1007/1-4020-2579-3_4}

\bibitem[\protect\citeauthoryear{{Elmegreen} \& {Scalo}}{{Elmegreen} \&
  {Scalo}}{2004}]{Elmegreen04}
{Elmegreen} B.~G.,  {Scalo} J.,  2004, \mn@doi [\araa]
  {10.1146/annurev.astro.41.011802.094859}, \href
  {https://ui.adsabs.harvard.edu/abs/2004ARA&A..42..211E} {42, 211}

\bibitem[\protect\citeauthoryear{{Ewen} \& {Purcell}}{{Ewen} \&
  {Purcell}}{1951}]{Ewen51}
{Ewen} H.~I.,  {Purcell} E.~M.,  1951, \mn@doi [\nat] {10.1038/168356a0}, \href
  {https://ui.adsabs.harvard.edu/abs/1951Natur.168..356E} {168, 356}

\bibitem[\protect\citeauthoryear{{Federrath}}{{Federrath}}{2018}]{Federrath18}
{Federrath} C.,  2018, \mn@doi [Physics Today] {10.1063/PT.3.3947}, \href
  {https://ui.adsabs.harvard.edu/abs/2018PhT....71f..38F} {71, 38}

\bibitem[\protect\citeauthoryear{{Federrath}, {Roman-Duval}, {Klessen},
  {Schmidt}  \& {Mac Low}}{{Federrath} et~al.}{2010}]{Federrath10}
{Federrath} C.,  {Roman-Duval} J.,  {Klessen} R.~S.,  {Schmidt} W.,   {Mac Low}
  M.~M.,  2010, \mn@doi [\aap] {10.1051/0004-6361/200912437}, \href
  {https://ui.adsabs.harvard.edu/abs/2010A&A...512A..81F} {512, A81}

\bibitem[\protect\citeauthoryear{{Federrath} et~al.,}{{Federrath}
  et~al.}{2017}]{Federrath17}
{Federrath} C.,  et~al., 2017, in {Crocker} R.~M.,  {Longmore} S.~N.,
  {Bicknell} G.~V.,  eds,  Vol. 322, The Multi-Messenger Astrophysics of the
  Galactic Centre. pp 123--128 (\mn@eprint {arXiv} {1609.08726}),
  \mn@doi{10.1017/S1743921316012357}

\bibitem[\protect\citeauthoryear{{Federrath}, {Klessen}, {Iapichino}  \&
  {Beattie}}{{Federrath} et~al.}{2021}]{Federrath21}
{Federrath} C.,  {Klessen} R.~S.,  {Iapichino} L.,   {Beattie} J.~R.,  2021,
  \mn@doi [Nature Astronomy] {10.1038/s41550-020-01282-z}, \href
  {https://ui.adsabs.harvard.edu/abs/2021NatAs...5..365F} {5, 365}

\bibitem[\protect\citeauthoryear{{Ferri{\`e}re}}{{Ferri{\`e}re}}{2001}]{Ferriere01}
{Ferri{\`e}re} K.~M.,  2001, \mn@doi [Reviews of Modern Physics]
  {10.1103/RevModPhys.73.1031}, \href
  {https://ui.adsabs.harvard.edu/abs/2001RvMP...73.1031F} {73, 1031}

\bibitem[\protect\citeauthoryear{{Ferri{\`e}re}}{{Ferri{\`e}re}}{2020}]{Ferriere2020}
{Ferri{\`e}re} K.,  2020, \mn@doi [Plasma Physics and Controlled Fusion]
  {10.1088/1361-6587/ab49eb}, \href
  {https://ui.adsabs.harvard.edu/abs/2020PPCF...62a4014F} {62, 014014}

\bibitem[\protect\citeauthoryear{{Field}}{{Field}}{1958}]{Field58}
{Field} G.~B.,  1958, \mn@doi [Proceedings of the IRE]
  {10.1109/JRPROC.1958.286741}, \href
  {https://ui.adsabs.harvard.edu/abs/1958PIRE...46..240F} {46, 240}

\bibitem[\protect\citeauthoryear{{Field}, {Goldsmith}  \& {Habing}}{{Field}
  et~al.}{1969}]{Field69}
{Field} G.~B.,  {Goldsmith} D.~W.,   {Habing} H.~J.,  1969, \mn@doi [\apjl]
  {10.1086/180324}, \href
  {https://ui.adsabs.harvard.edu/abs/1969ApJ...155L.149F} {155, L149}

\bibitem[\protect\citeauthoryear{{Fukui}, {Hayakawa}, {Inoue}, {Torii},
  {Okamoto}, {Tachihara}, {Onishi}  \& {Hayashi}}{{Fukui}
  et~al.}{2018}]{Fukui18}
{Fukui} Y.,  {Hayakawa} T.,  {Inoue} T.,  {Torii} K.,  {Okamoto} R.,
  {Tachihara} K.,  {Onishi} T.,   {Hayashi} K.,  2018, \mn@doi [\apj]
  {10.3847/1538-4357/aac16c}, \href
  {https://ui.adsabs.harvard.edu/abs/2018ApJ...860...33F} {860, 33}

\bibitem[\protect\citeauthoryear{{Gatto} et~al.,}{{Gatto}
  et~al.}{2015}]{Gatto15}
{Gatto} A.,  et~al., 2015, \mn@doi [\mnras] {10.1093/mnras/stv324}, \href
  {https://ui.adsabs.harvard.edu/abs/2015MNRAS.449.1057G} {449, 1057}

\bibitem[\protect\citeauthoryear{{Gazol} \& {Villagran}}{{Gazol} \&
  {Villagran}}{2016}]{Gazol16}
{Gazol} A.,  {Villagran} M.~A.,  2016, \mn@doi [\mnras]
  {10.1093/mnras/stw1789}, \href
  {https://ui.adsabs.harvard.edu/abs/2016MNRAS.462.2033G} {462, 2033}

\bibitem[\protect\citeauthoryear{{HI4PI Collaboration} et~al.,}{{HI4PI
  Collaboration} et~al.}{2016}]{HI4PI16}
{HI4PI Collaboration} et~al., 2016, \mn@doi [\aap]
  {10.1051/0004-6361/201629178}, \href
  {https://ui.adsabs.harvard.edu/abs/2016A&A...594A.116H} {594, A116}

\bibitem[\protect\citeauthoryear{{Hagen}, {Lilley}  \& {McClain}}{{Hagen}
  et~al.}{1955}]{Hagen55}
{Hagen} J.~P.,  {Lilley} A.~E.,   {McClain} E.~F.,  1955, \mn@doi [\apj]
  {10.1086/146096}, \href
  {https://ui.adsabs.harvard.edu/abs/1955ApJ...122..361H} {122, 361}

\bibitem[\protect\citeauthoryear{Harris et~al.,}{Harris
  et~al.}{2020}]{harris2020array}
Harris C.~R.,  et~al., 2020, \mn@doi [Nature] {10.1038/s41586-020-2649-2}, 585,
  357

\bibitem[\protect\citeauthoryear{{Haud}}{{Haud}}{2000}]{Haud00}
{Haud} U.,  2000, \aap, \href
  {https://ui.adsabs.harvard.edu/abs/2000A&A...364...83H} {364, 83}

\bibitem[\protect\citeauthoryear{{Haud} \& {Kalberla}}{{Haud} \&
  {Kalberla}}{2007}]{Haud07}
{Haud} U.,  {Kalberla} P.~M.~W.,  2007, \mn@doi [\aap]
  {10.1051/0004-6361:20065796}, \href
  {https://ui.adsabs.harvard.edu/abs/2007A&A...466..555H} {466, 555}

\bibitem[\protect\citeauthoryear{{Heiles} \& {Troland}}{{Heiles} \&
  {Troland}}{2003a}]{Heiles03}
{Heiles} C.,  {Troland} T.~H.,  2003a, \mn@doi [\apjs] {10.1086/367785}, \href
  {https://ui.adsabs.harvard.edu/abs/2003ApJS..145..329H} {145, 329}

\bibitem[\protect\citeauthoryear{{Heiles} \& {Troland}}{{Heiles} \&
  {Troland}}{2003b}]{Heiles03II}
{Heiles} C.,  {Troland} T.~H.,  2003b, \mn@doi [\apj] {10.1086/367828}, \href
  {https://ui.adsabs.harvard.edu/abs/2003ApJ...586.1067H} {586, 1067}

\bibitem[\protect\citeauthoryear{{Heiles} \& {Troland}}{{Heiles} \&
  {Troland}}{2005}]{Heiles05}
{Heiles} C.,  {Troland} T.~H.,  2005, \mn@doi [\apj] {10.1086/428896}, \href
  {https://ui.adsabs.harvard.edu/abs/2005ApJ...624..773H} {624, 773}

\bibitem[\protect\citeauthoryear{{Hennebelle} \& {P{\'e}rault}}{{Hennebelle} \&
  {P{\'e}rault}}{1999}]{Hennebelle99}
{Hennebelle} P.,  {P{\'e}rault} M.,  1999, \aap, \href
  {https://ui.adsabs.harvard.edu/abs/1999A&A...351..309H} {351, 309}

\bibitem[\protect\citeauthoryear{{Hu}, {Lazarian}, {Alina}, {Pogosyan}  \&
  {Ho}}{{Hu} et~al.}{2023}]{Hu23}
{Hu} Y.,  {Lazarian} A.,  {Alina} D.,  {Pogosyan} D.,   {Ho} K.~W.,  2023,
  \mn@doi [\mnras] {10.1093/mnras/stad1924}, \href
  {https://ui.adsabs.harvard.edu/abs/2023MNRAS.524.2994H} {524, 2994}

\bibitem[\protect\citeauthoryear{Hunter}{Hunter}{2007}]{Hunter:2007}
Hunter J.~D.,  2007, \mn@doi [Computing in Science \& Engineering]
  {10.1109/MCSE.2007.55}, 9, 90

\bibitem[\protect\citeauthoryear{{Jenkins} \& {Tripp}}{{Jenkins} \&
  {Tripp}}{2011}]{Jenkins11}
{Jenkins} E.~B.,  {Tripp} T.~M.,  2011, \mn@doi [\apj]
  {10.1088/0004-637X/734/1/65}, \href
  {https://ui.adsabs.harvard.edu/abs/2011ApJ...734...65J} {734, 65}

\bibitem[\protect\citeauthoryear{{Kalberla} \& {Haud}}{{Kalberla} \&
  {Haud}}{2018}]{Kalberla18}
{Kalberla} P.~M.~W.,  {Haud} U.,  2018, \mn@doi [\aap]
  {10.1051/0004-6361/201833146}, \href
  {https://ui.adsabs.harvard.edu/abs/2018A&A...619A..58K} {619, A58}

\bibitem[\protect\citeauthoryear{{Kalberla} \& {Kerp}}{{Kalberla} \&
  {Kerp}}{2009}]{Kalberla09}
{Kalberla} P. M.~W.,  {Kerp} J.,  2009, \mn@doi [\araa]
  {10.1146/annurev-astro-082708-101823}, \href
  {https://ui.adsabs.harvard.edu/abs/2009ARA&A..47...27K} {47, 27}

\bibitem[\protect\citeauthoryear{{Kanekar}, {Subrahmanyan}, {Chengalur}  \&
  {Safouris}}{{Kanekar} et~al.}{2003}]{Kanekar03}
{Kanekar} N.,  {Subrahmanyan} R.,  {Chengalur} J.~N.,   {Safouris} V.,  2003,
  \mn@doi [\mnras] {10.1111/j.1365-2966.2003.07333.x}, \href
  {https://ui.adsabs.harvard.edu/abs/2003MNRAS.346L..57K} {346, L57}

\bibitem[\protect\citeauthoryear{{Kanekar}, {Braun}  \& {Roy}}{{Kanekar}
  et~al.}{2011}]{Kanekar11}
{Kanekar} N.,  {Braun} R.,   {Roy} N.,  2011, \mn@doi [\apjl]
  {10.1088/2041-8205/737/2/L33}, \href
  {https://ui.adsabs.harvard.edu/abs/2011ApJ...737L..33K} {737, L33}

\bibitem[\protect\citeauthoryear{{Kim} \& {Ostriker}}{{Kim} \&
  {Ostriker}}{2017}]{Kim17}
{Kim} C.-G.,  {Ostriker} E.~C.,  2017, \mn@doi [\apj]
  {10.3847/1538-4357/aa8599}, \href
  {https://ui.adsabs.harvard.edu/abs/2017ApJ...846..133K} {846, 133}

\bibitem[\protect\citeauthoryear{{Kim}, {Ostriker}  \& {Kim}}{{Kim}
  et~al.}{2013}]{Kim13}
{Kim} C.-G.,  {Ostriker} E.~C.,   {Kim} W.-T.,  2013, \mn@doi [\apj]
  {10.1088/0004-637X/776/1/1}, \href
  {https://ui.adsabs.harvard.edu/abs/2013ApJ...776....1K} {776, 1}

\bibitem[\protect\citeauthoryear{{Kim}, {Ostriker}  \& {Kim}}{{Kim}
  et~al.}{2014}]{Kim14}
{Kim} C.-G.,  {Ostriker} E.~C.,   {Kim} W.-T.,  2014, \mn@doi [\apj]
  {10.1088/0004-637X/786/1/64}, \href
  {https://ui.adsabs.harvard.edu/abs/2014ApJ...786...64K} {786, 64}

\bibitem[\protect\citeauthoryear{{Kim}, {Kim}, {Gong}  \& {Ostriker}}{{Kim}
  et~al.}{2023}]{Kim23}
{Kim} C.-G.,  {Kim} J.-G.,  {Gong} M.,   {Ostriker} E.~C.,  2023, \mn@doi
  [\apj] {10.3847/1538-4357/acbd3a}, \href
  {https://ui.adsabs.harvard.edu/abs/2023ApJ...946....3K} {946, 3}

\bibitem[\protect\citeauthoryear{{Koley} \& {Roy}}{{Koley} \&
  {Roy}}{2019}]{KR19}
{Koley} A.,  {Roy} N.,  2019, \mn@doi [\mnras] {10.1093/mnras/sty3152}, \href
  {https://ui.adsabs.harvard.edu/abs/2019MNRAS.483..593K} {483, 593}

\bibitem[\protect\citeauthoryear{{Koyama} \& {Inutsuka}}{{Koyama} \&
  {Inutsuka}}{2000}]{Koyama00}
{Koyama} H.,  {Inutsuka} S.-I.,  2000, \mn@doi [\apj] {10.1086/308594}, \href
  {https://ui.adsabs.harvard.edu/abs/2000ApJ...532..980K} {532, 980}

\bibitem[\protect\citeauthoryear{{Koyama} \& {Inutsuka}}{{Koyama} \&
  {Inutsuka}}{2002}]{Koyama02}
{Koyama} H.,  {Inutsuka} S.-i.,  2002, \mn@doi [\apjl] {10.1086/338978}, \href
  {https://ui.adsabs.harvard.edu/abs/2002ApJ...564L..97K} {564, L97}

\bibitem[\protect\citeauthoryear{{Kriel}, {Beattie}, {Seta}  \&
  {Federrath}}{{Kriel} et~al.}{2022}]{Kriel22}
{Kriel} N.,  {Beattie} J.~R.,  {Seta} A.,   {Federrath} C.,  2022, \mn@doi
  [\mnras] {10.1093/mnras/stac969}, \href
  {https://ui.adsabs.harvard.edu/abs/2022MNRAS.513.2457K} {513, 2457}

\bibitem[\protect\citeauthoryear{{Kulkarni} \& {Heiles}}{{Kulkarni} \&
  {Heiles}}{1988}]{Kulkarni88}
{Kulkarni} S.~R.,  {Heiles} C.,  1988, in {Kellermann} K.~I.,  {Verschuur}
  G.~L.,  eds, , Galactic and Extragalactic Radio Astronomy.
pp 95--153

\bibitem[\protect\citeauthoryear{{Larson}}{{Larson}}{1979}]{Larson79}
{Larson} R.~B.,  1979, \mn@doi [\mnras] {10.1093/mnras/186.3.479}, \href
  {https://ui.adsabs.harvard.edu/abs/1979MNRAS.186..479L} {186, 479}

\bibitem[\protect\citeauthoryear{{Larson}}{{Larson}}{1981}]{Larson81}
{Larson} R.~B.,  1981, \mn@doi [\mnras] {10.1093/mnras/194.4.809}, \href
  {https://ui.adsabs.harvard.edu/abs/1981MNRAS.194..809L} {194, 809}

\bibitem[\protect\citeauthoryear{{Lazarian} \& {Pogosyan}}{{Lazarian} \&
  {Pogosyan}}{2000}]{Lazarian20}
{Lazarian} A.,  {Pogosyan} D.,  2000, \mn@doi [\apj] {10.1086/309040}, \href
  {https://ui.adsabs.harvard.edu/abs/2000ApJ...537..720L} {537, 720}

\bibitem[\protect\citeauthoryear{{Lei} \& {Clark}}{{Lei} \&
  {Clark}}{2022}]{Lei22}
{Lei} M.,  {Clark} S.~E.,  2022, \mn@doi [arXiv e-prints]
  {10.48550/arXiv.2212.06182}, \href
  {https://ui.adsabs.harvard.edu/abs/2022arXiv221206182L} {p. arXiv:2212.06182}

\bibitem[\protect\citeauthoryear{{Lindner} et~al.,}{{Lindner}
  et~al.}{2015}]{Lindner15}
{Lindner} R.~R.,  et~al., 2015, \mn@doi [\aj] {10.1088/0004-6256/149/4/138},
  \href {https://ui.adsabs.harvard.edu/abs/2015AJ....149..138L} {149, 138}

\bibitem[\protect\citeauthoryear{{Liszt}}{{Liszt}}{2001}]{Liszt01}
{Liszt} H.,  2001, \mn@doi [\aap] {10.1051/0004-6361:20010395}, \href
  {https://ui.adsabs.harvard.edu/abs/2001A&A...371..698L} {371, 698}

\bibitem[\protect\citeauthoryear{{Mac Low} \& {Klessen}}{{Mac Low} \&
  {Klessen}}{2004}]{MacLow04}
{Mac Low} M.-M.,  {Klessen} R.~S.,  2004, \mn@doi [Reviews of Modern Physics]
  {10.1103/RevModPhys.76.125}, \href
  {https://ui.adsabs.harvard.edu/abs/2004RvMP...76..125M} {76, 125}

\bibitem[\protect\citeauthoryear{{Marchal} \&
  {Miville-Desch{\^e}nes}}{{Marchal} \&
  {Miville-Desch{\^e}nes}}{2021}]{Marchal21}
{Marchal} A.,  {Miville-Desch{\^e}nes} M.-A.,  2021, \mn@doi [\apj]
  {10.3847/1538-4357/abd108}, \href
  {https://ui.adsabs.harvard.edu/abs/2021ApJ...908..186M} {908, 186}

\bibitem[\protect\citeauthoryear{{Marchal}, {Miville-Desch{\^e}nes}, {Orieux},
  {Gac}, {Soussen}, {Lesot}, {d'Allonnes}  \& {Salom{\'e}}}{{Marchal}
  et~al.}{2019}]{Marchal19}
{Marchal} A.,  {Miville-Desch{\^e}nes} M.-A.,  {Orieux} F.,  {Gac} N.,
  {Soussen} C.,  {Lesot} M.-J.,  {d'Allonnes} A.~R.,   {Salom{\'e}} Q.,  2019,
  \mn@doi [\aap] {10.1051/0004-6361/201935335}, \href
  {https://ui.adsabs.harvard.edu/abs/2019A&A...626A.101M} {626, A101}

\bibitem[\protect\citeauthoryear{{Martin}, {Blagrave}, {Lockman}, {Pinheiro
  Gon{\c{c}}alves}, {Boothroyd}, {Joncas}, {Miville-Desch{\^e}nes}  \&
  {Stephan}}{{Martin} et~al.}{2015}]{Martin15}
{Martin} P.~G.,  {Blagrave} K.~P.~M.,  {Lockman} F.~J.,  {Pinheiro
  Gon{\c{c}}alves} D.,  {Boothroyd} A.~I.,  {Joncas} G.,
  {Miville-Desch{\^e}nes} M.~A.,   {Stephan} G.,  2015, \mn@doi [\apj]
  {10.1088/0004-637X/809/2/153}, \href
  {https://ui.adsabs.harvard.edu/abs/2015ApJ...809..153M} {809, 153}

\bibitem[\protect\citeauthoryear{McClure-Griffiths, Stanimirovi\'{c}  \&
  Rybarczyk}{McClure-Griffiths et~al.}{2023}]{Griffiths23}
McClure-Griffiths N.~M.,  Stanimirovi\'{c} S.,   Rybarczyk D.~R.,  2023,
  \mn@doi [Annual Review of Astronomy and Astrophysics]
  {10.1146/annurev-astro-052920-104851}, 61, null

\bibitem[\protect\citeauthoryear{{McKee}}{{McKee}}{1995}]{McKee95}
{McKee} C.~F.,  1995, in {Ferrara} A.,  {McKee} C.~F.,  {Heiles} C.,
  {Shapiro} P.~R.,  eds,  Astronomical Society of the Pacific Conference Series
  Vol. 80, The Physics of the Interstellar Medium and Intergalactic Medium.
  p.~292

\bibitem[\protect\citeauthoryear{{McKee} \& {Ostriker}}{{McKee} \&
  {Ostriker}}{1977}]{McKee77}
{McKee} C.~F.,  {Ostriker} J.~P.,  1977, \mn@doi [\apj] {10.1086/155667}, \href
  {https://ui.adsabs.harvard.edu/abs/1977ApJ...218..148M} {218, 148}

\bibitem[\protect\citeauthoryear{{McKee} \& {Ostriker}}{{McKee} \&
  {Ostriker}}{2007}]{McKee07}
{McKee} C.~F.,  {Ostriker} E.~C.,  2007, \mn@doi [\araa]
  {10.1146/annurev.astro.45.051806.110602}, \href
  {https://ui.adsabs.harvard.edu/abs/2007ARA&A..45..565M} {45, 565}

\bibitem[\protect\citeauthoryear{{Miville-Desch{\^e}nes}, {Levrier}  \&
  {Falgarone}}{{Miville-Desch{\^e}nes} et~al.}{2003}]{MD03}
{Miville-Desch{\^e}nes} M.~A.,  {Levrier} F.,   {Falgarone} E.,  2003, \mn@doi
  [\apj] {10.1086/376603}, \href
  {https://ui.adsabs.harvard.edu/abs/2003ApJ...593..831M} {593, 831}

\bibitem[\protect\citeauthoryear{{Miville-Desch{\^e}nes}
  et~al.,}{{Miville-Desch{\^e}nes} et~al.}{2017}]{MD17}
{Miville-Desch{\^e}nes} M.~A.,  et~al., 2017, \mn@doi [\aap]
  {10.1051/0004-6361/201628289}, \href
  {https://ui.adsabs.harvard.edu/abs/2017A&A...599A.109M} {599, A109}

\bibitem[\protect\citeauthoryear{{Muller} \& {Oort}}{{Muller} \&
  {Oort}}{1951}]{Muller51}
{Muller} C.~A.,  {Oort} J.~H.,  1951, \mn@doi [\nat] {10.1038/168357a0}, \href
  {https://ui.adsabs.harvard.edu/abs/1951Natur.168..357M} {168, 357}

\bibitem[\protect\citeauthoryear{{Murray} et~al.,}{{Murray}
  et~al.}{2015}]{Murray15}
{Murray} C.~E.,  et~al., 2015, \mn@doi [\apj] {10.1088/0004-637X/804/2/89},
  \href {https://ui.adsabs.harvard.edu/abs/2015ApJ...804...89M} {804, 89}

\bibitem[\protect\citeauthoryear{{Murray}, {Stanimirovi{\'c}}, {Kim},
  {Ostriker}, {Lindner}, {Heiles}, {Dickey}  \& {Babler}}{{Murray}
  et~al.}{2017}]{Murray17}
{Murray} C.~E.,  {Stanimirovi{\'c}} S.,  {Kim} C.-G.,  {Ostriker} E.~C.,
  {Lindner} R.~R.,  {Heiles} C.,  {Dickey} J.~M.,   {Babler} B.,  2017, \mn@doi
  [\apj] {10.3847/1538-4357/aa5d12}, \href
  {https://ui.adsabs.harvard.edu/abs/2017ApJ...837...55M} {837, 55}

\bibitem[\protect\citeauthoryear{{Murray}, {Stanimirovi{\'c}}, {Goss},
  {Heiles}, {Dickey}, {Babler}  \& {Kim}}{{Murray} et~al.}{2018}]{Murray18}
{Murray} C.~E.,  {Stanimirovi{\'c}} S.,  {Goss} W.~M.,  {Heiles} C.,  {Dickey}
  J.~M.,  {Babler} B.,   {Kim} C.-G.,  2018, \mn@doi [\apjs]
  {10.3847/1538-4365/aad81a}, \href
  {https://ui.adsabs.harvard.edu/abs/2018ApJS..238...14M} {238, 14}

\bibitem[\protect\citeauthoryear{{Murray}, {Stanimirovi{\'c}}, {Heiles},
  {Dickey}, {McClure-Griffiths}, {Lee}, {M. Goss}  \&
  {Killerby-Smith}}{{Murray} et~al.}{2021}]{Murray21}
{Murray} C.~E.,  {Stanimirovi{\'c}} S.,  {Heiles} C.,  {Dickey} J.~M.,
  {McClure-Griffiths} N.~M.,  {Lee} M.~Y.,  {M. Goss} W.,   {Killerby-Smith}
  N.,  2021, \mn@doi [\apjs] {10.3847/1538-4365/ac0f0b}, \href
  {https://ui.adsabs.harvard.edu/abs/2021ApJS..256...37M} {256, 37}

\bibitem[\protect\citeauthoryear{Newville et~al.,}{Newville
  et~al.}{2023}]{lmfit23}
Newville M.,  et~al., 2023, lmfit/lmfit-py: 1.2.1,
  \mn@doi{10.5281/zenodo.7887568}, \url
  {https://doi.org/10.5281/zenodo.7887568}

\bibitem[\protect\citeauthoryear{{Nguyen}, {Dawson}, {Lee}, {Murray},
  {Stanimirovi{\'c}}, {Heiles}, {Miville-Desch{\^e}nes}  \& {Petzler}}{{Nguyen}
  et~al.}{2019}]{Nguyen19}
{Nguyen} H.,  {Dawson} J.~R.,  {Lee} M.-Y.,  {Murray} C.~E.,
  {Stanimirovi{\'c}} S.,  {Heiles} C.,  {Miville-Desch{\^e}nes} M.~A.,
  {Petzler} A.,  2019, \mn@doi [\apj] {10.3847/1538-4357/ab2b9f}, \href
  {https://ui.adsabs.harvard.edu/abs/2019ApJ...880..141N} {880, 141}

\bibitem[\protect\citeauthoryear{{Padoan}, {Federrath}, {Chabrier}, {Evans},
  {Johnstone}, {J{\o}rgensen}, {McKee}  \& {Nordlund}}{{Padoan}
  et~al.}{2014}]{Padoan14}
{Padoan} P.,  {Federrath} C.,  {Chabrier} G.,  {Evans} N.~J. I.,  {Johnstone}
  D.,  {J{\o}rgensen} J.~K.,  {McKee} C.~F.,   {Nordlund} {\r{A}}.,  2014, in
  {Beuther} H.,  {Klessen} R.~S.,  {Dullemond} C.~P.,   {Henning} T.,  eds,
  Protostars and Planets VI. pp 77--100 (\mn@eprint {arXiv} {1312.5365}),
  \mn@doi{10.2458/azu_uapress_9780816531240-ch004}

\bibitem[\protect\citeauthoryear{{Rathjen} et~al.,}{{Rathjen}
  et~al.}{2021}]{Rathjen21}
{Rathjen} T.-E.,  et~al., 2021, \mn@doi [\mnras] {10.1093/mnras/stab900}, \href
  {https://ui.adsabs.harvard.edu/abs/2021MNRAS.504.1039R} {504, 1039}

\bibitem[\protect\citeauthoryear{{Roy}, {Peedikakkandy}  \& {Chengalur}}{{Roy}
  et~al.}{2008}]{Roy08}
{Roy} N.,  {Peedikakkandy} L.,   {Chengalur} J.~N.,  2008, \mn@doi [\mnras]
  {10.1111/j.1745-3933.2008.00473.x}, \href
  {https://ui.adsabs.harvard.edu/abs/2008MNRAS.387L..18R} {387, L18}

\bibitem[\protect\citeauthoryear{{Roy}, {Kanekar}, {Braun}  \&
  {Chengalur}}{{Roy} et~al.}{2013a}]{Roy13}
{Roy} N.,  {Kanekar} N.,  {Braun} R.,   {Chengalur} J.~N.,  2013a, \mn@doi
  [\mnras] {10.1093/mnras/stt1743}, \href
  {https://ui.adsabs.harvard.edu/abs/2013MNRAS.436.2352R} {436, 2352}

\bibitem[\protect\citeauthoryear{{Roy}, {Kanekar}  \& {Chengalur}}{{Roy}
  et~al.}{2013b}]{Roy13II}
{Roy} N.,  {Kanekar} N.,   {Chengalur} J.~N.,  2013b, \mn@doi [\mnras]
  {10.1093/mnras/stt1746}, \href
  {https://ui.adsabs.harvard.edu/abs/2013MNRAS.436.2366R} {436, 2366}

\bibitem[\protect\citeauthoryear{{Saha}, {Roy}  \& {Bhattacharya}}{{Saha}
  et~al.}{2018}]{Saha18}
{Saha} P.,  {Roy} N.,   {Bhattacharya} M.,  2018, \mn@doi [\mnras]
  {10.1093/mnrasl/sly139}, \href
  {https://ui.adsabs.harvard.edu/abs/2018MNRAS.480L.126S} {480, L126}

\bibitem[\protect\citeauthoryear{{Saury}, {Miville-Desch{\^e}nes},
  {Hennebelle}, {Audit}  \& {Schmidt}}{{Saury} et~al.}{2014}]{Saury14}
{Saury} E.,  {Miville-Desch{\^e}nes} M.~A.,  {Hennebelle} P.,  {Audit} E.,
  {Schmidt} W.,  2014, \mn@doi [\aap] {10.1051/0004-6361/201321113}, \href
  {https://ui.adsabs.harvard.edu/abs/2014A&A...567A..16S} {567, A16}

\bibitem[\protect\citeauthoryear{{Scalo} \& {Elmegreen}}{{Scalo} \&
  {Elmegreen}}{2004}]{Scalo04}
{Scalo} J.,  {Elmegreen} B.~G.,  2004, \mn@doi [\araa]
  {10.1146/annurev.astro.42.120403.143327}, \href
  {https://ui.adsabs.harvard.edu/abs/2004ARA&A..42..275S} {42, 275}

\bibitem[\protect\citeauthoryear{{Seifried}, {Schmidt}  \&
  {Niemeyer}}{{Seifried} et~al.}{2011}]{Seifried11}
{Seifried} D.,  {Schmidt} W.,   {Niemeyer} J.~C.,  2011, \mn@doi [\aap]
  {10.1051/0004-6361/201014373}, \href
  {https://ui.adsabs.harvard.edu/abs/2011A&A...526A..14S} {526, A14}

\bibitem[\protect\citeauthoryear{{Seon} \& {Kim}}{{Seon} \&
  {Kim}}{2020}]{Seon20}
{Seon} K.-i.,  {Kim} C.-G.,  2020, \mn@doi [\apjs] {10.3847/1538-4365/aba2d6},
  \href {https://ui.adsabs.harvard.edu/abs/2020ApJS..250....9S} {250, 9}

\bibitem[\protect\citeauthoryear{{Seta} \& {Federrath}}{{Seta} \&
  {Federrath}}{2022}]{Seta22}
{Seta} A.,  {Federrath} C.,  2022, \mn@doi [\mnras] {10.1093/mnras/stac1400},
  \href {https://ui.adsabs.harvard.edu/abs/2022MNRAS.514..957S} {514, 957}

\bibitem[\protect\citeauthoryear{{Sharma}, {Parrish}  \& {Quataert}}{{Sharma}
  et~al.}{2010}]{Sharma2010}
{Sharma} P.,  {Parrish} I.~J.,   {Quataert} E.,  2010, \mn@doi [\apj]
  {10.1088/0004-637X/720/1/652}, \href
  {https://ui.adsabs.harvard.edu/abs/2010ApJ...720..652S} {720, 652}

\bibitem[\protect\citeauthoryear{{Strasser} \& {Taylor}}{{Strasser} \&
  {Taylor}}{2004}]{Strasser04}
{Strasser} S.,  {Taylor} A.~R.,  2004, \mn@doi [\apj] {10.1086/381674}, \href
  {https://ui.adsabs.harvard.edu/abs/2004ApJ...603..560S} {603, 560}

\bibitem[\protect\citeauthoryear{{Thomson} et~al.,}{{Thomson}
  et~al.}{2019}]{Thomson19}
{Thomson} A. J.~M.,  et~al., 2019, \mn@doi [\mnras] {10.1093/mnras/stz1438},
  \href {https://ui.adsabs.harvard.edu/abs/2019MNRAS.487.4751T} {487, 4751}

\bibitem[\protect\citeauthoryear{{V{\'a}zquez-Semadeni}, {G{\'o}mez},
  {Jappsen}, {Ballesteros-Paredes}, {Gonz{\'a}lez}  \&
  {Klessen}}{{V{\'a}zquez-Semadeni} et~al.}{2007}]{Vazquez-SemadeniEA2007}
{V{\'a}zquez-Semadeni} E.,  {G{\'o}mez} G.~C.,  {Jappsen} A.~K.,
  {Ballesteros-Paredes} J.,  {Gonz{\'a}lez} R.~F.,   {Klessen} R.~S.,  2007,
  \mn@doi [\apj] {10.1086/510771}, \href
  {https://ui.adsabs.harvard.edu/abs/2007ApJ...657..870V} {657, 870}

\bibitem[\protect\citeauthoryear{Virtanen et~al.,}{Virtanen
  et~al.}{2020}]{2020SciPy-NMeth}
Virtanen P.,  et~al., 2020, \mn@doi [Nature Methods]
  {10.1038/s41592-019-0686-2}, \href {https://rdcu.be/b08Wh} {17, 261}

\bibitem[\protect\citeauthoryear{{Wakker} \& {van Woerden}}{{Wakker} \& {van
  Woerden}}{1997}]{Wakker97}
{Wakker} B.~P.,  {van Woerden} H.,  1997, \mn@doi [\araa]
  {10.1146/annurev.astro.35.1.217}, \href
  {https://ui.adsabs.harvard.edu/abs/1997ARA&A..35..217W} {35, 217}

\bibitem[\protect\citeauthoryear{{Wolfire}, {Hollenbach}, {McKee}, {Tielens}
  \& {Bakes}}{{Wolfire} et~al.}{1995}]{Wolfire95}
{Wolfire} M.~G.,  {Hollenbach} D.,  {McKee} C.~F.,  {Tielens} A.~G.~G.~M.,
  {Bakes} E.~L.~O.,  1995, \mn@doi [\apj] {10.1086/175510}, \href
  {https://ui.adsabs.harvard.edu/abs/1995ApJ...443..152W} {443, 152}

\bibitem[\protect\citeauthoryear{{Wolfire}, {McKee}, {Hollenbach}  \&
  {Tielens}}{{Wolfire} et~al.}{2003}]{Wolfire03}
{Wolfire} M.~G.,  {McKee} C.~F.,  {Hollenbach} D.,   {Tielens} A.~G.~G.~M.,
  2003, \mn@doi [\apj] {10.1086/368016}, \href
  {https://ui.adsabs.harvard.edu/abs/2003ApJ...587..278W} {587, 278}

\bibitem[\protect\citeauthoryear{{Wouthuysen}}{{Wouthuysen}}{1952}]{Wouthuysen52}
{Wouthuysen} S.~A.,  1952, \mn@doi [\aj] {10.1086/106661}, \href
  {https://ui.adsabs.harvard.edu/abs/1952AJ.....57R..31W} {57, 31}

\bibitem[\protect\citeauthoryear{pandas~development team}{pandas~development
  team}{2020}]{reback2020pandas}
pandas~development team T.,  2020, pandas-dev/pandas: Pandas,
  \mn@doi{10.5281/zenodo.3509134}, \url
  {https://doi.org/10.5281/zenodo.3509134}

\makeatother
\end{thebibliography}
%%%%%%%%%%%%%%%%%%%%%%%%%%%%%%%%%%%%%%%%%%%%%%%%%%

%%%%%%%%%%%%%%%%% APPENDICES %%%%%%%%%%%%%%%%%%%%%

\appendix

\section{Gaussian Decomposition Algorithm}\label{appendix:gauss_decomp_algo}

\subsection{Fitting single spectrum}
Here, we describe our method to decompose a single absorption spectrum into multiple Gaussian components:
\begin{itemize}
    \item We estimate the number of Gaussian components by two methods. The first estimation is equal to the number of local maxima in the spectra, which we call $N_{\rm max}$. The second estimation is from the number of changes in curvature (second derivative) of the spectrum, which can pick up unresolved Gaussians. We call this $N_{curv}$. It is easy to see that $N_{curv}\geq N_{\rm max}$. The positions of the peaks or the curvature change and their corresponding heights give the estimates for the mean and amplitude of the Gaussians. The fits are performed by allowing a variation of $\pm5\mathrm{\ km~s^{-1}}$ around the calculated mean estimates and no restrictions on width or amplitude.

    \item We perform two separate incremental fits to the spectrum. At each step, we increase the number of Gaussians by one and check the corresponding $\mathrm{reduced-}\chi^2$ value. As the spectra are noiseless, $\mathrm{red-}\chi^2$ here simply means $\frac{\mathrm{Least\ Sq.}}{\mathrm{Deg.\ of\ Freedom}}$. The first incremental fit is from $N_{\rm max}$ to $N_{curv}$ and the second from $N_{curv}$ to some suitable limit to the number of Gaussian. We take the upper limit to be $N_{curve}+10$. For the additional components above $N_{curve}$, we use initial parameter values of $0\mathrm{\ km~s^{-1}}$, $1\mathrm{\ km~s^{-1}}$ and $1$ for the mean, $\sigma$, and amplitude respectively, and place no restrictions on any of the parameter values.

    \item For each incremental fit, we terminate the increment if one of the following occurs, as they indicate the onset of overfitting: 1) The $\mathrm{red-}\chi^2$ value does not decrease by at least 10\% 2) the peak of any component is higher than the highest value of the spectrum 3) the width of any component is narrower than $0.1\mathrm{\ km~s^{-1}}$ 4) the centers and width of any two components are within $0.5\mathrm{\ km~s^{-1}}$ of each other.

    \item From the two incremental fits, we take the fit with the least $\mathrm{red-}\chi^2$ value.
\end{itemize}

For the spectra with noise (for the analysis described in Appendix \ref{appendix:noise_effect}), we slightly modify the method. We describe the method below:

\begin{itemize}
    \item We apply the Savitzky-Golay filter (with a smoothening window of $2\mathrm{\ km~s^{-1}}$ and a polynomial order of three) to smoothen the spectrum. Following this, we use the peak identification algorithm by \emph{scipy} to identify the peaks above a threshold prominence level of $0.05$. We only find peaks separated by at least $1.2\mathrm{\ km~s^{-1}}$. We call the number of peaks as $N_{peak}$. As computing the curvature of noisy spectra is difficult, we do not perform this step.

    \item We perform an incremental fit from $N_{peak}$ to $N_{peak}+10$ and at each step check the corresponding red-$\chi^2$ value. This time, red-$\chi^2$ is defined as $\frac{\mathrm{Least\ Sq.}}{\mathrm{Deg.\ of\ Freedom}\times \sigma_{\tau}^2}$, where $\sigma_\tau$ is the noise level in absorption spectrum. 

    \item The termination checks are the same as in the previous case, except for the first point. The modified first termination check is: the red-$\chi^2$ value does not move further away from unity. 
\end{itemize}

We note that even after applying the termination checks, there can be over or under-fitting of the spectra. This, however, is inevitable, as the total spectrum may not exactly be a sum of Gaussians. Also, Gaussian decomposition is not a well-posed problem in general, as Gaussians do not form an orthonormal basis \citep{Haud07,Lindner15}. 

\subsection{Joint decomposition}

For the joint decomposition, we first fit the absorption spectrum (method described in the previous section). We use the absorption spectrum Gaussian components to fit the emission spectrum by restricting the Gaussian center and width (allowing a variation of $\pm0.5\mathrm{\ km~s^{-1}}$) and allow for up to $5$ additional components to account for any warm components missed in the absorption decomposition, though we do not restrict the additional Gaussian components' width to be in the warm regime. For the additional components, we use initial parameter values of $0\mathrm{\ km~s^{-1}}$, $1\mathrm{\ km~s^{-1}}$ and $1\mathrm{\ K}$ for the mean, $\sigma$, and amplitude respectively, and we place no restrictions on any of the parameter values. The termination in fitting is done following the termination criteria discussed in the previous section. 

\subsection{Component acceptance criteria}\label{appendix:filtering_comps}

\subsubsection{JGD}

The maximum temperature associated with a component is given by $T_{\rm max}^{\rm comp} = 121\sigma_v^2$. Due to non-thermal broadening, the kinetic temperature of the component $T_k^{\rm comp}$ is usually less than $T_{\rm max}^{\rm comp}$, and $T_s^{\rm comp}$ is, in turn, less than or equal to $T_k^{\rm comp}$ because of insufficient collisions or radiative coupling. Thus, a required physical condition for the Gaussian components is $T_s^{\rm comp}<T_{\rm max}^{\rm comp}$. In our decomposition, we do not force this condition while performing the fit to allow the algorithm to span a larger parameter space, which was seen to improve the quality of the fit, but for most components, the condition is seen to be satisfied naturally, which is also an indication that the Gaussian decomposition can extract the true physical Gaussian components. For the constant WF effect case, $\gtrapprox 80\%$ of the components were found to have $T_s^{\rm comp}<T_{\rm max}^{\rm comp}$, though for the maximum WF effect case (see \S\ref{subsec:mwf_discussion}), only around $66\%$ of components have $T_s^{\rm comp}<T_{\rm max}^{\rm comp}$. For the latter, for several components, we expect $T_s^{\rm comp}\approx T_{\rm max}^{\rm comp}$ when turbulence broadening is relatively small. Uncertainty in Gaussian decomposition and the related error may lead to some components having slightly higher estimated spin temperature than the maximum temperature. But in this case, too, we see that $\approx 80\%$ of the component satisfies the relation $T_s^{\rm comp}<1.1\times T_{\rm max}^{\rm comp}$. Thus for most components, the spin temperature lies very close to the limit of $T_{\rm max}^{\rm comp}$. Such occurrences of a small fraction of components with $T_s^{\rm comp}>T_{\rm max}^{\rm comp}$ have also been noted with observed data \citep{Heiles03}, even with careful decomposition of the data. These arise either due to misfitting of the spectra or due to incorrectly assigned Gaussian counterparts in absorption and emission spectra. For our analysis, in both cases, we eliminate all components for which $T_s^{\rm comp}>1.1\times T_{\rm max}^{\rm comp}$, with the $10\%$ error window chosen to allow for errors related to Gaussian fitting, even for the true components.  Additionally, we also discard all components with $T_B^p<0.1$K and $\tau^p<5\times10^{-4}$K. 

\subsubsection{KR Method}

The KR method considers that a fraction $f$ of the total broadening is of non-thermal origin and iteratively solves for this fraction (for details, see \citealt{KR19}). We impose the convergence that two consecutive values of $f$ being within $5\%$ of each other within $200$ iterations. We disregard all the components for which convergence is not attained. We also discard all components for which the inferred kinetic temperature is $>15000\mathrm{\ K}$. Additionally, we also discard all components $\tau^p<5\times10^{-4}$K. We briefly note that these criteria yield a higher number of absorption components than the JGD criteria described in the previous section. However, no significant difference in the final inferences is seen if the JGD criteria-inferred absorption components are used for the KR method. 

We note that the termination criteria for Gaussian fitting and subsequent component selection criteria were chosen mostly based on testing the algorithm on a few spectra. Small changes in the chosen values of the associated parameters were seen not to affect the results much, even quantitatively. 

\section{The effect of noise}\label{appendix:noise_effect}

\begin{figure}
    \centering
    \includegraphics[width=\linewidth]{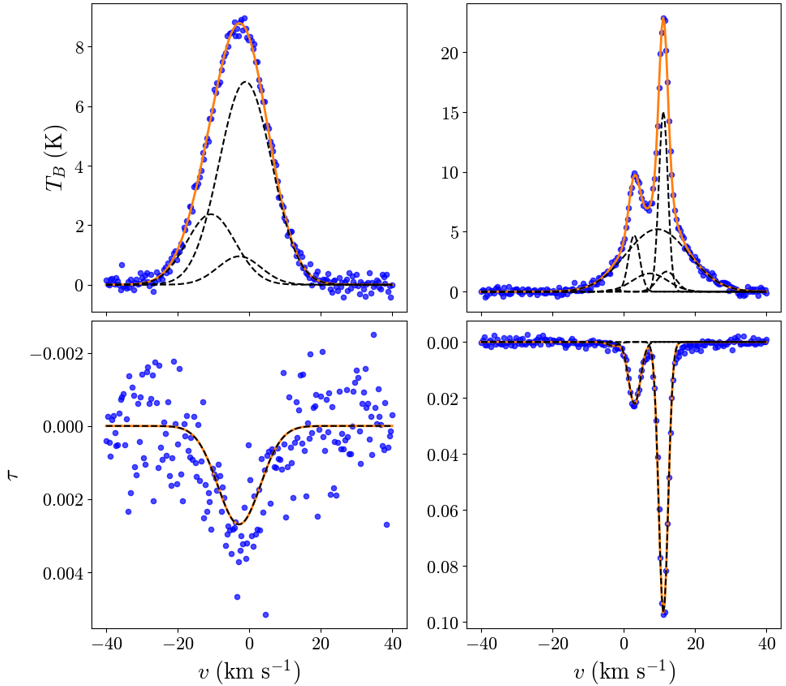}
    \caption{Examples of emission and absorption spectra with noise and the corresponding Gaussian components inferred using the JGD method.}
    \label{fig:example_spectrum_with_noise}
\end{figure}

\begin{figure}
    \centering
    \includegraphics[width=\linewidth]{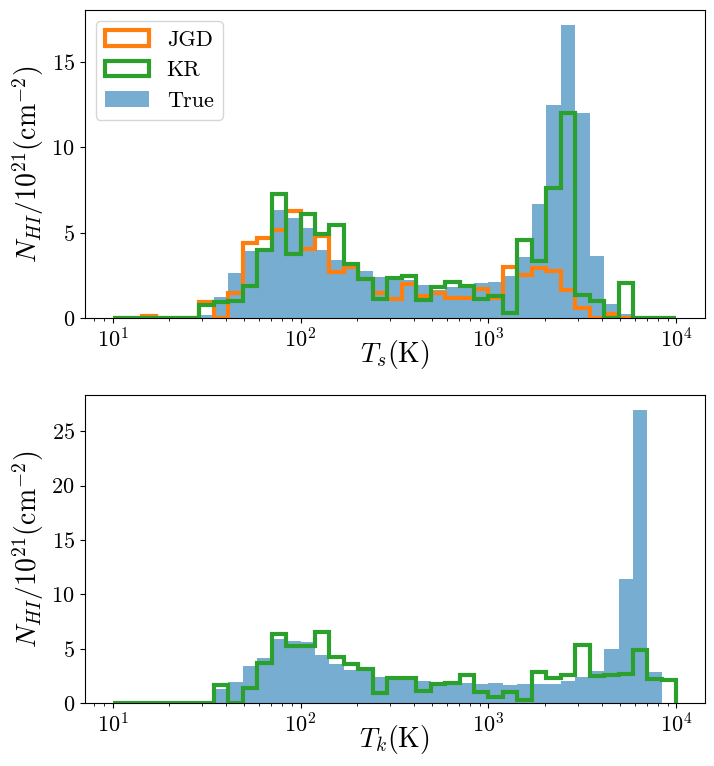}
    \caption{Inferred column density distribution from spectra with noise (for figure details, see caption of Figure \ref{fig:nh_dist_cwf_jgd_cwf}). The inferred distribution agrees well with the true distribution in the lower temperature regime. The underestimation at higher temperatures is more prominent than with the noiseless spectra.}
    \label{fig:nh_dist_cwf_jgd_cwf_with_noise}
\end{figure}

\begin{figure}
    \centering
    \includegraphics[width=\linewidth]{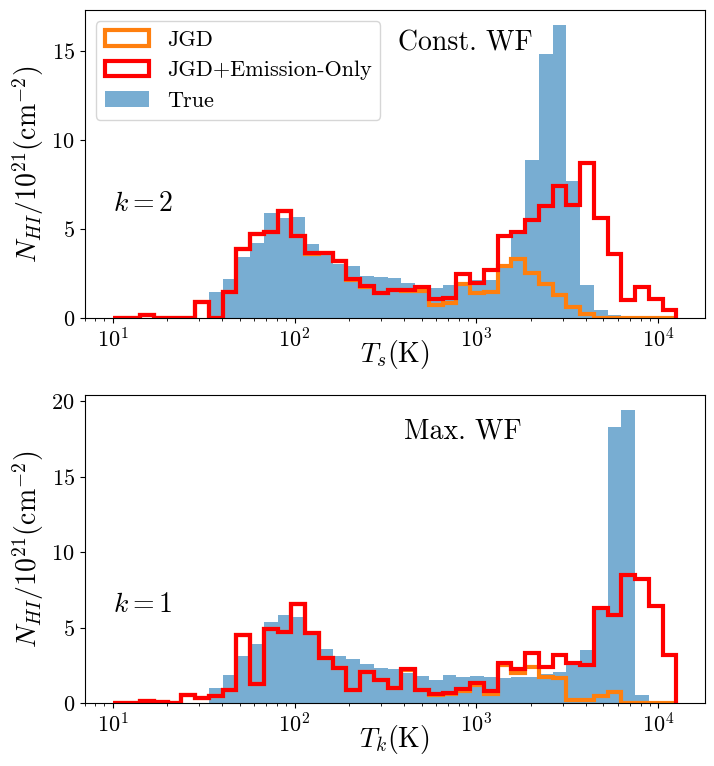}
    \caption{Inferred column density distribution with JGD with and without the inclusion of the emission-only components. The spin temperature of the emission-only components is estimated using Equation \ref{eq:est_em_only_ts}. For reasonable values of $k$, the difference mostly lies in the warm gas regime.}
    \label{fig:nh_dist_cwf_jgd_cwf_with_noise_with_emonly_0.002}
\end{figure}

Noise in observed ISM spectra can potentially affect the analyses and the inferences. To test this, we add noise to the synthetic spectra (generated using the methodology described in \S\ref{subsec:spectrum_generation}) and perform our analyses. For the optical depth and emission spectra, we add random Gaussian noise with standard deviations of $\sigma_{\tau}=10^{-3}$ and $\sigma_{T_B}=0.2\mathrm{\ K}$ respectively in each velocity channel, following the treatment in \citet{Murray18}. These values are in ballpark agreement with the noise levels in other surveys. We perform this analysis with the $512^3$ simulation.

The $T_B-\tau$ distribution is not affected by noise except at the low optical depth/brightness temperature regime of the warm gas. The values of the parameters $T_{B,w}$ and $C$ (see Eq. \ref{eq:tnw_equation}) do not change significantly. However, the estimation of parameter $T_{w}$ becomes difficult with higher noise levels. High-sensitivity absorption surveys \citep[like][where $\sigma_{\tau}\approx2\times10^{-4}$]{Roy13}, which can yield a sufficient number of $>3\sigma$ spectral data points in the low optical depth regime of $\tau\lesssim0.005$, are needed for a proper estimation of this model parameter.

We also test the Gaussian decomposition analysis methods. Gaussian decomposition of noisy spectra is challenging, and several dedicated algorithms have been devised for this task \citep{Lindner15,Marchal19}. In this work, we employ a much simpler method and perform a similar incremental Gaussian fitting as with the noiseless spectra (with appropriate modifications; see Appendix \ref{appendix:gauss_decomp_algo} for the details). Figure \ref{fig:example_spectrum_with_noise} shows a couple of examples of spectra with noise and the inferred Gaussian components. We perform the same analyses as done in \S\ref{subsec:using_both_abs_em} and \S\ref{subsec:using_only abs} with the same $200$ lines of sight we used previously. 

Figure \ref{fig:nh_dist_cwf_jgd_cwf_with_noise} shows the inferred column density distribution for the constant WF effect case. Even with noise, the recovered distributions with JGD and KR methods agree well with the true distribution in the lower temperature regime of $T_S/T_k\lessapprox2500\mathrm{\ K}$. This is mostly because the optical depth and brightness temperature values of these components are significantly higher than the noise level. Due to noise, however, the low optical depth warm components are further suppressed in the absorption spectra compared to noiseless spectra. This manifests both as the lower number of inferred absorption components ($390$ compared to $546$ for noiseless spectra in constant WF effect case) and an increased underestimation of the warm gas column density from the noiseless spectra. The results are similar for the maximum WF effect case. See Table \ref{tab:gas_fractions} for the summary of the results.

We check how the inclusion of the emission-only components affects the column density distribution over temperature. As these components are undetected in absorption, we assume that the peak optical depth of these components is close to the noise level, $\sigma_{\tau}$. The spin temperature can be estimated from the following relation:
\begin{equation}\label{eq:est_em_only_ts}
    T_{s}^{\rm comp,\rm em}=\frac{T_B^{\rm est,peak}}{k\sigma_{\tau}}
\end{equation}
where $T_B^{\rm est,peak}$ is the estimated peak brightness temperature of the warm components (see Equation \ref{eq:tb_true_est}) and $k$ is a suitably chosen factor. Figure \ref{fig:nh_dist_cwf_jgd_cwf_with_noise_with_emonly_0.002} shows the revised column density distribution over spin temperature for both the constant and maximum WF effect cases (top and bottom respectively, red histogram) when the emission-only components are included with the JGD analysis. For each case, we use the value of $k$, which makes the resultant distribution closer to the actual distribution. For comparison, we also show the distribution without the emission-only components (orange histogram). As expected, the difference lies mostly in the warm region. We clearly see that by including the emission-only components, we recover significant amounts of the warm gas column density.

\section{The effect of simulation resolution}\label{app:resolution_effect}

\begin{table}
\centering
\caption{The phase fractions in terms of mass (volume) for the three different resolutions. We use the definition of gas phases in terms of $T_k$ as defined in \S\ref{sec:simulations} for the simulation domains with different resolutions.}
\label{tab:nh_frac_128_252}
\resizebox{0.9\linewidth}{!}{%
\begin{tabular}{|c|c|c|c|}
\hline 
    & $f_\mathrm{CNM}$        & $f_\mathrm{UNM}$        & $f_\mathrm{WNM}$         \\
    \hline
$128$ & 0.16 (0.01) & 0.58 (0.43) & 0.26 (0.56) \\
$252$ & 0.25 (0.01) & 0.44 (0.29) & 0.31 (0.70)  \\
$512$ & 0.32 (0.02) & 0.34 (0.23) & 0.33 (0.75)  \\
\hline
\end{tabular}
}
\end{table}

\begin{figure}
    \centering
    \includegraphics[width=\linewidth]{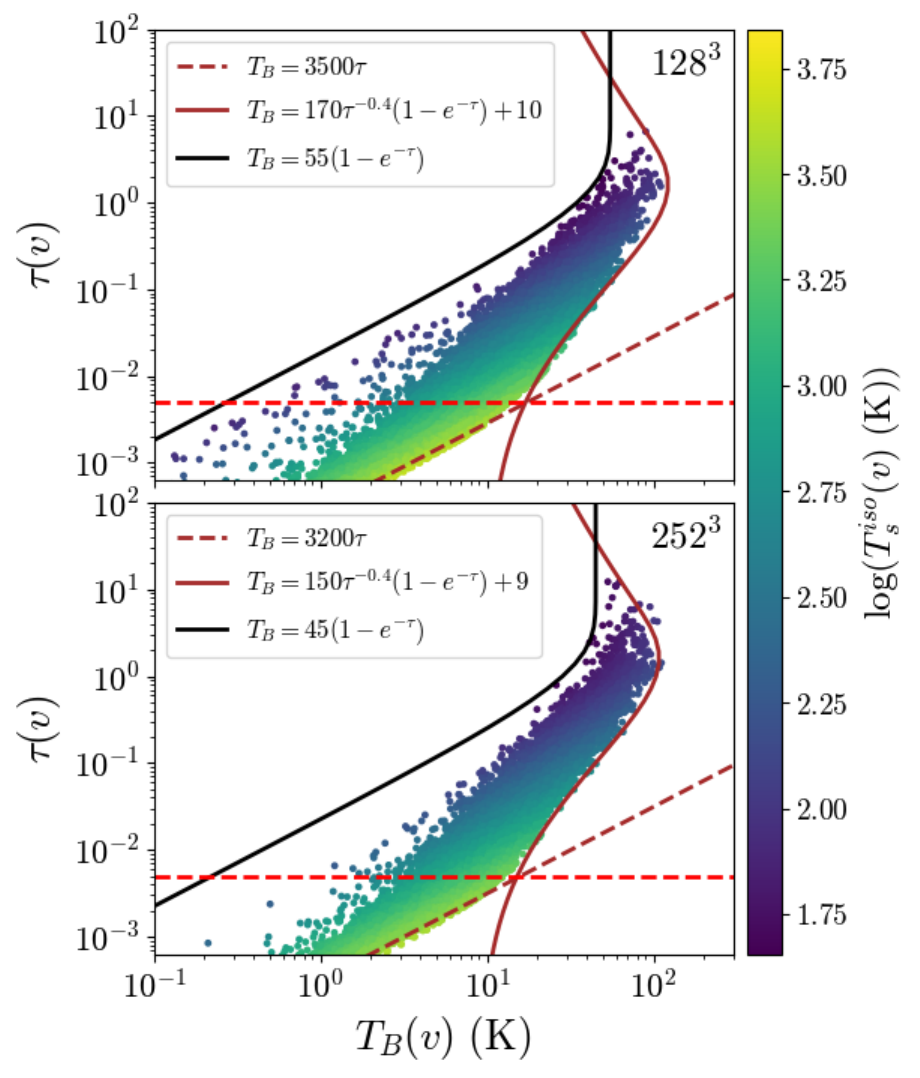}
    \caption{The $T_B(v)-\tau(v)$ distribution for $128^3$ (\textbf{Top}) and $252^3$ (\textbf{Bottom}) simulations for randomly chosen $1000$ lines of sight. For the $512^3$, see Figure~\ref{fig:tb_tau_inner_both} (left panel).}
    \label{fig:tb_tau_128_252}
\end{figure}

\begin{figure}
    \centering
    \includegraphics[width=\linewidth]{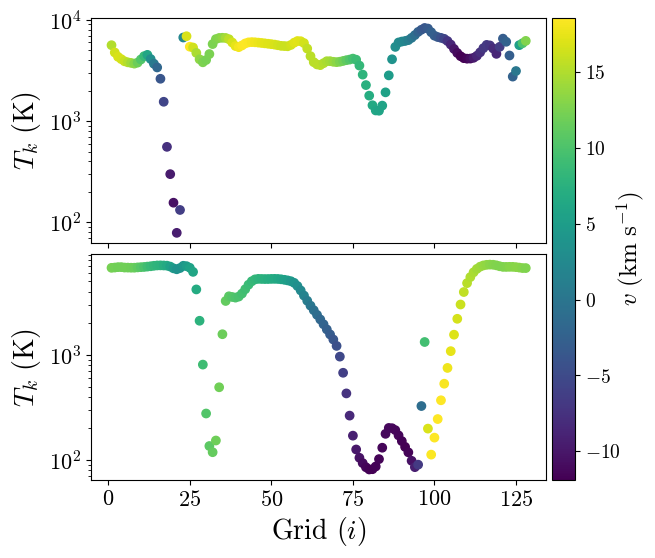}
    \caption{Examples of discontinuities in physical parameters of adjacent grid cells (both in temperature and velocity in the top figure at $\approx23\mathrm{\ km~s^{-1}}$, and in velocity in the bottom figure at $\approx95\mathrm{\ km~s^{-1}}$) arising from numerical effects, which leads to the data points in the left region of the $T_B(v)-\tau(v)$ distribution. In ISM, however, similar effects can be expected with fast-moving cold clouds through the ambient medium (see \S\ref{subsubsec:left_tbtau_models}).}
    \label{fig:128_problem_exs}
\end{figure}

\begin{figure*}
    \centering
    \includegraphics[width=\linewidth]{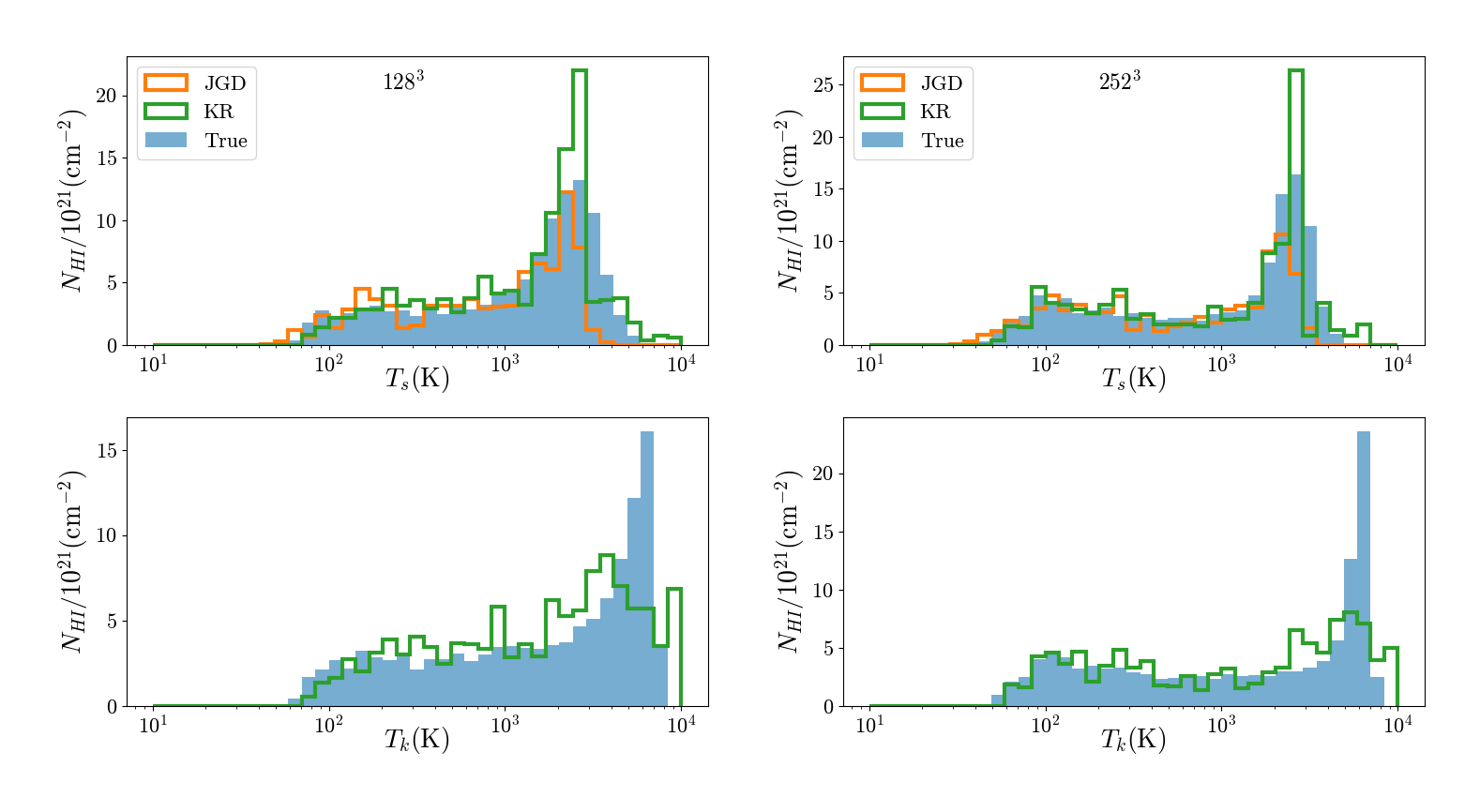}
    \caption{The inferred column density distribution for $128^3$ (\textbf{Left}) and $252^3$ (\textbf{Right}) simulations for randomly chosen $200$ lines of sight using both JGD and KR methods. The non-warm distribution is well recovered. This shows that the analysis methods are stable with resolution. For the distributions with $512^3$ grids simulation, see Figure \ref{fig:nh_dist_cwf_jgd_cwf}.}
    \label{fig:nh_dist_128_252}
\end{figure*}

As a convergence test, we used simulations with different resolutions, namely $128^3$ and $252^3$. The amount of cold, unstable, and warm gas in the medium is dependent on the resolution, with the fraction of unstable gas in the medium decreasing with increasing resolution (see Table \ref{tab:nh_frac_128_252} for the mass and volume fractions of the gas phases). This is because simulations with lower resolution fail to adequately resolve the cooling fronts at scales comparable to the cooling length, which mostly occur on unresolved scales. We perform the analysis only for noiseless spectra.

The difference in cold and unstable gas morphology in the lower resolution affects the $T_B(v)-\tau(v)$ distribution of the spectral data. Figure \ref{fig:tb_tau_128_252} shows the distribution and the model for the boundaries along with the fit parameters for the constant WF effect case. The value of the parameter $C$ (see Eq. \ref{eq:tnw_equation}) is higher than the value obtained for the $512^3$ simulation data, while the other two parameter values do not differ significantly. For the maximum WF effect case, the parameter $T_w$ takes the value of $\approx6000\mathrm{\ K}$, leaving the other parameters nearly unchanged from the constant WF effect distribution. The lowest resolution ($128^3$ grid) simulation domain contains a significant number of data points in the left region, which decreases with increasing resolution. The origin of the data points in the lowest resolution domain has been identified to be cold clouds with velocity and temperature distinctly different from its immediate boundaries (see Figure \ref{fig:128_problem_exs}. This is an artifact of insufficient resolution, where the transition between surrounding gas components is often not well resolved. However, a similar effect may be achieved in the ISM with fast-moving cold clouds (see \S\ref{subsubsec:left_tbtau_models}). The frequency of such cases, and consequently the number of data points in the left, decreases with increasing resolution. Broadly, the $T_B(v)-\tau(v)$ distribution and the model parameters also converge with increasing resolution. 

We apply the Gaussian decomposition methods (both JGD and KR methods) to the spectra generated from the lower-resolution simulation of $200$ randomly chosen lines of sight. Figure \ref{fig:nh_dist_128_252} shows the results for the same. Similar to all previous cases, the column density distribution of the non-warm gas is recovered well. The distribution is significantly different for the $128^3$ simulation, however, the distribution converges with increasing simulation resolution.

\section{Effect of emission beam width}\label{appendix:emission_beam_effect}

\begin{figure}
    \centering
    \includegraphics[width=\linewidth]{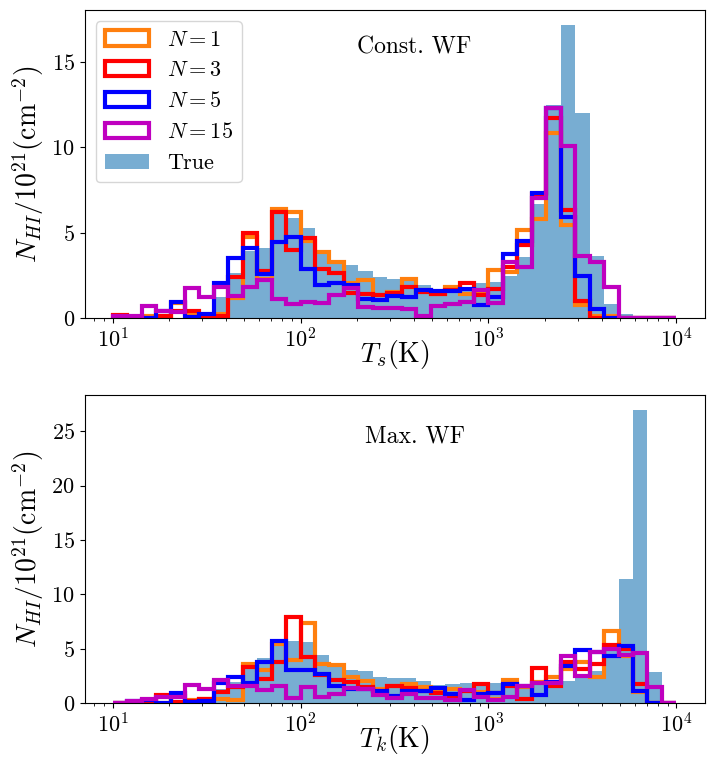}
    \caption{The inferred column density distribution using JGD when the emission spectra are averaged over $N\times N$ pixels around the absorption pixel for $N=3$ (red), $N=5$ (blue), and $N=15$ (violet) for both the constant and maximum WF effect cases with the $512^3$ grids simulation. For comparison, the inferred distribution with no averaging ($N=1$) is provided (orange).}
    \label{fig:jgd_3c3_5c5}
\end{figure}

\begin{figure}
    \centering
    \includegraphics[width=\linewidth]{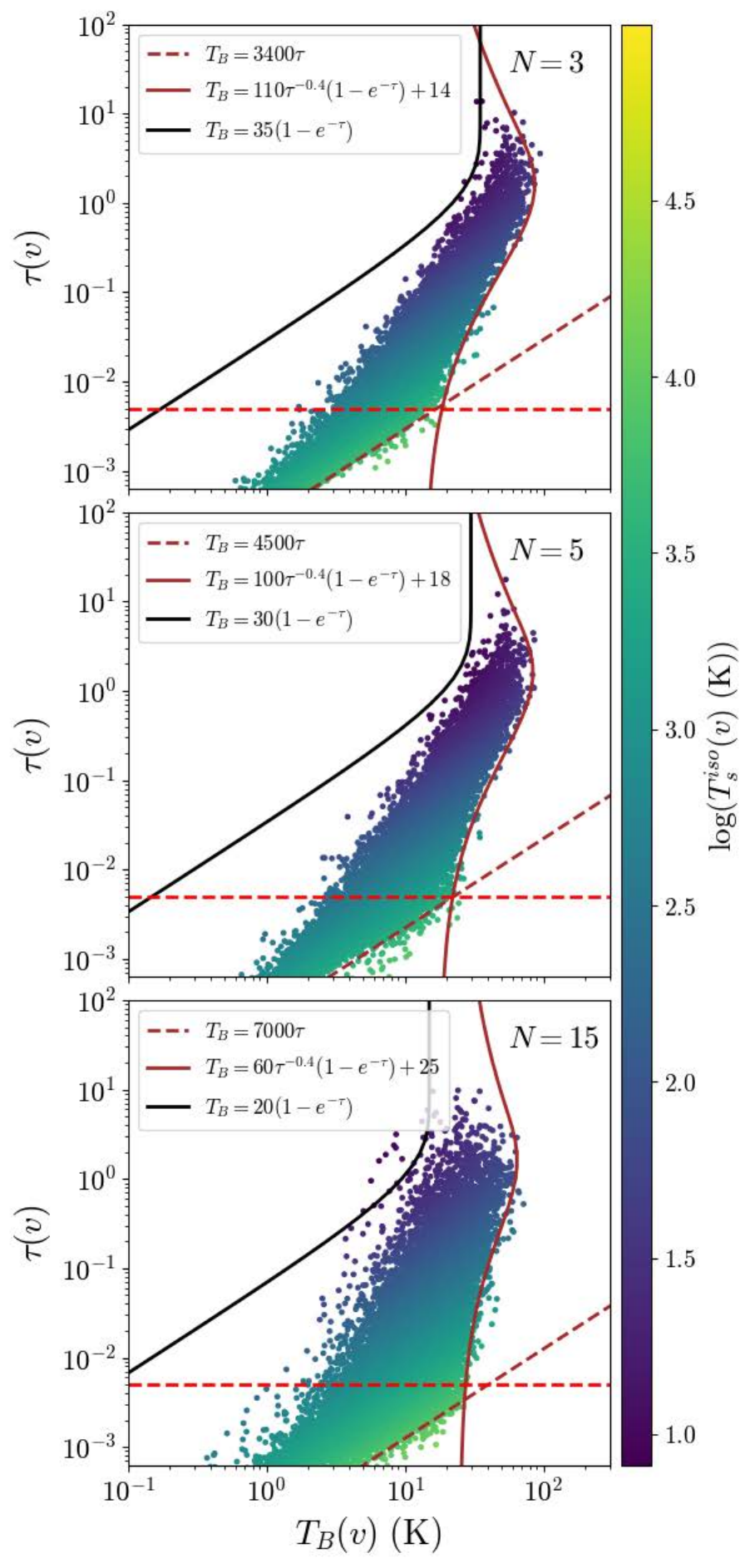}
    \caption{The $T_B(v)-\tau(v)$ distribution with emission spectrum averaged over $N\times N$ pixels around the absorption pixel for $N=3$ (top), $N=5$ (middle), and $N=15$ (bottom) for randomly chosen $1000$ lines of sight with the $512^3$ grids simulation.}
    \label{fig:N_3_5_tb_tau}
\end{figure}

As already mentioned in \S\ref{subsubsec:lim_of_obs_analy_techs}, the emission spectra are recorded over a much larger area in the sky than absorption spectra. We test how this may affect the inferences and the outcomes of the analysis techniques by averaging the emission spectrum over a region of $NXN$ around the pixel used to generate the absorption spectrum. \textcolor{black}{Here we consider three cases of $N=3,\ 5,\ \text{and}\ 15$, which correspond to $\sim1.2\mathrm{\ pc},\ \sim2\mathrm{\ pc},\ \text{and}\ \sim6\mathrm{\ pc}$ in spatial scale. We note here that the case of $N=15$ has been included to demonstrate how very broad emission beams may affect the inferences. Averaging over such a spatial scale smears all the smaller-scale cold gas structures (most of which have scales of $1-10\mathrm{\ pc}$). However, this may not happen in observations, and the cold structures are recovered well in spectra.}

To substantiate our argument regarding the $N=15$ case, we do a rough estimate of the averaging that may happen in observations. The high-resolution emission surveys like HI4PI or Arecibo, which are used in such studies, have beam widths of $3-4'$. The furthest ends of the Milky Way correspond to a distance of $\sim10\mathrm{\ kpc}$. This amounts to a spatial scale of $\sim10\mathrm{\ pc}$. Thus, this corresponds to the largest possible scale of averaging (at the farthest ends of the galaxy). Thus, averaging over $\sim6\mathrm{\ pc}$ scale throughout for a $200\mathrm{\ pc}$ simulation domain is an extreme case, with which we try to magnify the effect (which also shows the strength of these numerical experiments). However, we recognize the possibility of similar spatial averaging effects in past surveys with large emission beam widths (e.g., LAB). This shows the importance of a systematic study utilizing surveys with different beam widths to constrain this effect (see~\S\ref{subsubsec:lim_of_obs_analy_techs} for a discussion). Such an analysis is beyond the scope of this paper, but we plan to do it in the future (see Appendix~\ref{app:lab_hi4pi}). Additionally, we note that the various parameters (for joint fitting and termination, see Appendix~\ref{appendix:gauss_decomp_algo}) used in our Gaussian decomposition algorithm may require changes, as the current version may not be optimized for such a large emission width. However, for a fair comparison, we have used the same parameters and algorithm throughout.

All parameters of the model for $T_B(v)-\tau(v)$ are affected when the emission beam size is increased, with the effect increasing with increasing $N$. Figure \ref{fig:N_3_5_tb_tau} shows the distributions for the two cases, and the model fits for the constant WF effect case. For the right boundary model, the parameter $C$ decreases, and the other two parameters increase. The left boundary also shifts towards the left (leading to a decrease in parameter $C'$). These translate to an overall decrease (increase) in the non-warm (warm) gas brightness temperatures. However, similar to all previous cases, only the parameter $T_w$ changes from constant to maximum WF effect case ($6000\mathrm{\ K}$, $7800\mathrm{\ K}$ and $15000\mathrm{\ K}$ for $N=3$, $N=5$ and $N=15$ respectively).

Figure \ref{fig:jgd_3c3_5c5} shows the inferred column density using JGD for the same $200$ lines of sight used in our work previously. The KR method does not depend on the emission spectra; thus, we do not include it. For comparison, we also show the JGD inference with the standard case of single-pixel emission spectra ($N=1$). We see that with increasing emission beam size, there is a tendency to underestimate the non-warm gas, with the underestimation being severe for $N=15$. No significant change in the warm gas estimation is seen. The quality of the joint Gaussian fits is seen to decrease with increasing $N$.  Overall, in all cases, a qualitative two-peak distribution is still recovered with a minimum in the unstable phase, though, for $N=15$, the inferred distribution differs significantly from the true distribution and shows a somewhat flat distribution throughout the cold and unstable phases.

The reason behind the above effects possibly lies in the fact that the cooler gas components have smaller scales. Averaging over a larger region reduces the significance of these components in the spectra. The properties of the warm gas, having larger scales, do not change much over the beam width, and thus, their contribution to the spectra is not significantly affected.

\section{Comparing LAB and HI4PI Spectra}\label{app:lab_hi4pi}

\begin{figure}
    \centering
    \includegraphics[width=\linewidth]{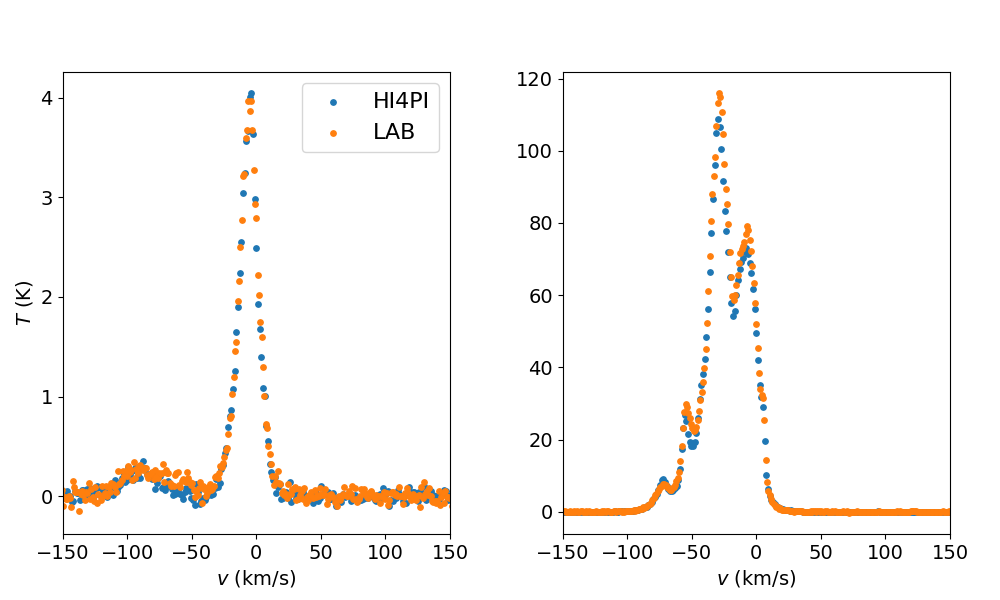}
    \caption{LAB and HI4PI emission data over-plotted for two representative lines of sight. Despite the different emission beam widths, the spectra are quite similar. Thus, we do not expect our inferences to change significantly if HI4PI data is included.}
    \label{fig:lab_hi4pi}
\end{figure}

As discussed at several places throughout the text (for example, see \S\ref{subsubsec:lim_of_obs_analy_techs} and Appendix \ref{appendix:emission_beam_effect}), emission spectra from surveys with narrow emission beam widths are important in constraining several ISM properties, especially in resolving the smaller-scale cold clouds. In \S\ref{subsec:tb_tau_model}, we have applied our $T_B-\tau$ distribution model to the data from \citet{Roy13}, where the emission data have been derived from the Leiden–Argentine–Bonn (LAB) survey. This survey suffers from having a large emission beam width of $\sim36'$. A possible way out is to use emission data from surveys with lower beam width, for example, the HI4PI survey \citep{HI4PI16}, whose data is available publicly. However, the major associated difficulty is to obtain the corresponding absorption spectra. A careful comparison and interpolation of the available absorption spectra from nearby lines of sight has to be performed, with appropriate consideration of the different spectral resolutions (for example, $1.5\mathrm{\ km~s^{-1}}$ resolution for HI4PI against $\approx0.3-0.5\mathrm{\ km~s^{-1}}$ for the corresponding absorption spectra from GMRT/WSRT/ATCA) and the noise levels. Otherwise, there may be spurious data points from the artifacts of interpolation that would be difficult to explain. This is beyond the scope of this paper, and for this work, we stick to the data from \citet{Roy13}. We note here that Roy et al.~are working towards a full data release of 60 lines of sight where a thorough comparison of results using HI4PI and LAB/GASS will be presented. We plan to re-do the analysis of this paper with the new data release. 

In order to assess how our results might change with the inclusion of HI4PI data, we compare the emission spectra of HI4PI and LAB for the 34~LAB lines of sight. Figure~\ref{fig:lab_hi4pi} presents the spectra for a couple of lines of sight as representatives of the sample. We see that the two emission spectra are very similar. Thus, we do not expect our results to change significantly if HI4PI data are included.

%%%%%%%%%%%%%%%%%%%%%%%%%%%%%%%%%%%%%%%%%%%%%%%%%%

% Don't change these lines
\bsp	% typesetting comment
\label{lastpage}
\end{document}